%% file: deepPHY_V11.tex
\documentclass[journal,comsoc,10pt]{IEEEtran}
\input{./input.tex}

\begin{document}
\title{DeepWiPHY: Deep Learning-based Receiver Design and Dataset for IEEE 802.11ax Systems}
\author{Yi Zhang, Akash Doshi, Rob Liston, Wai-tian Tan, Xiaoqing Zhu, 
	\\Jeffrey G. Andrews, \IEEEmembership{Fellow,~IEEE}, and Robert W. Heath Jr., \IEEEmembership{Fellow,~IEEE}
	\thanks{Manuscript received April 02, 2020; revised August 19, 2020; accepted October 12, 2020. Date of publication xx xx, 2020; date of current version xx xx, 2020. The work of Y. Zhang and R. Heath at The University of Texas at Austin was supported in part by the Army Research Office under grant W911NF1910221 and the National Science Foundation under grant CNS-1731658. The associate editor coordinating the review of this article and approving it for publication was Dr. Deniz Gunduz. 
	\newline \indent Yi Zhang, Akash Doshi, Jeffrey G. Andrews, and Robert W. Heath Jr. are with The University of Texas at Austin, TX, 78731, US (e-mail: yi.zhang.cn@utexas.edu; akashsdoshi@utexas.edu; jandrews@ece.utexas.edu; rheath@utexas.edu).
	\newline \indent Rob Liston, Wai-tian Tan, and Xiaoqing Zhu are with Cisco System Inc., US (e-mail: rliston@cisco.com; dtan2@cisco.com; xiaoqzhu@cisco.com).
	}}
\markboth{}
{}
\maketitle

\begin{abstract}
In this work, we develop DeepWiPHY, a deep learning-based architecture to replace the channel estimation, common phase error (CPE) correction,
sampling rate offset (SRO) correction, and equalization modules of 
IEEE 802.11ax based orthogonal frequency division multiplexing (OFDM) receivers.
We first train DeepWiPHY with a synthetic dataset, which is generated
using representative indoor channel models and includes 
typical radio frequency (RF) impairments that are the source of nonlinearity in wireless systems. 
To further train and evaluate DeepWiPHY with real-world data,
we develop a passive sniffing-based data collection testbed composed of
Universal Software Radio Peripherals (USRPs) and commercially available IEEE 802.11ax products. 
The comprehensive evaluation of DeepWiPHY with synthetic and
real-world datasets (110 million synthetic OFDM symbols and 14 million real-world OFDM symbols) confirms that, even without fine-tuning the neural network's architecture parameters, 
DeepWiPHY achieves comparable performance to or outperforms the conventional WLAN receivers, 
in terms of both bit error rate (BER) and packet error rate (PER),
under a wide range of channel models, signal-to-noise (SNR) levels, and modulation schemes.
\end{abstract}


\section{Introduction}\label{sec:Intro}
\subsection{Motivation}\label{subsec:Motivation} 
\lettrine{D}{eep learning} (DL) has recently begun impacting modern wireless communications through 
their innate ability to recognize hidden patterns and correlations in raw data~\cite{Qin_DL_PHY_Overview_2019}.
Great engineering efforts have been devoted to optimizing individual components 
in the wireless physical (PHY) layer for the past decades.
In contrast, DL provides an end-to-end optimization methodology,
which can potentially prolong the ongoing wireless evolution~\cite{OShea_PHY_DL_2017}.

Though wireless communications is a maturing field, 
numerous DL applications in wireless systems have shown
potential room for improvement.
For example, a DL-based automatic modulation recognizer 
was proposed in~\cite{DL_Modulation_Recognition_GuanGui}.
Its high identification accuracy enables the practical implementation of 
cognitive radio systems. 
In~\cite{DeepCodeKim}, a recurrent neural network (RNN)-based code,
"Deepcode", was proposed for a statistically well-defined feedback channel, 
and beats state-of-the-art codes by three orders of magnitude. 
Two convolutional neural network (CNN) based autoencoders were built in~\cite{Zhu_transceiver_optimization_NN_2019,CNN_End2End_arbitrary_block_length}. 
The input of those autoencoders is a bit sequence of arbitrary length and their experiments empirically demonstrated the superiority of autoencoders compared to conventional transceivers. 

Designing a DL-based baseband receiver
could prove beneficial for the following reasons: 
(i) it can jointly optimize channel estimation and equalization
modules which greatly impact the eventual wireless PHY layer performance, 
and (ii) it has the potential to learn and compensate for the accumulated
nonlinear effects and estimation error in previous modules.

\subsection{Challenges}\label{subsec:Challenges} 
Even though the above examples have shown the potential of DL 
in wireless systems, exploiting DL at the PHY layer,
especially for DL-based baseband receiver designs, 
is still challenging due to the three reasons as detailed below:

\textbf{The difficulty in conceiving a more sophisticated redesign of conventional PHY layer implementations}: Conventional algorithms are optimal or near-optimal for most
individual components at the wireless PHY layer. This limits the room for improvement when a single component such as channel estimation, equalization, 
or detection is considered. Some researchers proposed to use autoencoders 
so that global optimality may be achieved and the nonlinear hardware impairments 
could be tackled all at once~\cite{OShea_PHY_DL_2017,DL_com_over_the_air_2018}. 
Unfortunately, as pointed out by~\cite{DL_com_over_the_air_2018},
it is nontrivial to train an autoencoder when radio frequency (RF) impairments are considered. 
Moreover, the unknown and dynamic wireless channel is also a challenge
for the neural network (NN) training. Therefore, it is still unclear if there exist any opportunities 
to make wireless PHY layer better with DL. Even if there were, would this gain be substantial?

\textbf{The trade-off between generalization and goal-specific development}: DL is a data-driven tool that is goal-specific. 
This means that a well trained NN can work 
only for a specific purpose. Existing wireless systems, however, strictly 
follows specific standards and protocols such as LTE~\cite{3GPP_V16} 
for 4G/5G cellular and IEEE 802.11ax~\cite{IEEE80211ax} for Wi-Fi 6. These imply that 
it is unrealistic to train a single DL-based receiver that can work for different standards.
Moreover, as DL is data-driven, any performance gain achieved without following
specific standards may not be valid for real wireless systems. Accordingly, it is still not clear how to incorporate DL into 
the existing standards, whether DL-based receivers could generalize well and 
what would be the corresponding performance improvement.

\textbf{The lack of relevant/real-world training data}: For a DL-based application, the quality and 
quantity of the dataset used for training sometimes could matter more than 
the NN architecture design itself~\cite{FewThingsOnML}. Unfortunately, 
unlike the original DL applications such as image classification and 
speech data analysis, getting relevant wireless datasets is much more challenging. 
First, conventional PHY layer designs are mostly model-driven and 
assume mutual independence, which accepts the use of synthetic data. 
But DL is a data-driven technology, and may result in poor 
generalization if only synthetic data is used for training.
Second, the wireless channel is highly environment-dependent and 
the transmitted wireless signal is designed and generated by
strictly following certain standards. 
These challenges suggest that a good training dataset should 
not only cover diverse channel scenarios but also follow specific standards.
Until now, there are no available 
open-access datasets and thus the real over-the-air performance of
DL-based receivers are still unknown~\cite{Qin_DL_PHY_Overview_2019}. 

Motivated by these observations, in this work, we focus on the joint
optimization of multiple wireless PHY layer components in 
orthogonal frequency division multiplexing (OFDM) receivers.
This involves channel estimation, common phase error (CPE) correction,
sampling rate offset (SRO) correction, and equalization modules. 
Specifically, we propose DeepWiPHY, a deep neural network (DNN) based architecture,
to replace the aforementioned conventional modules.
In particular, we design and train DeepWiPHY strictly following 
the latest lower band WLAN standard, i.e., \textbf{IEEE 802.11ax}. 
To experimentally train and evaluate DeepWiPHY, 
we further develop a \textbf{real-world data collection} framework 
using state-of-the-art IEEE 802.11ax products.
DeepWiPHY focuses on single-spatial-stream mode, 
which can be potentially extended to the multiple-spatial-streams mode in the future. 

\subsection{Contributions}\label{subsec:contribution} 
Our contributions are threefold: we design a DL-based receiver module, generate standard-based synthetic and real-world datasets, and perform a comprehensive practical evaluation of our proposed receiver using the generated data. We now describe the three contributions in detail.

	\textbf{DL-based receiver design}: We propose DeepWiPHY, a DL-based aggregated module that jointly 
	performs channel estimation, common phase error and sampling rate offset (CpeSro)
	correction, and equalization. In comparison with the conventional receiver
	which is highly modularized and suboptimal, DeepWiPHY performs 
	multiple compensations all at once, which can potentially achieve better performance 
	and also deal with the nonlinear RF impairments.
	In contrast with autoencoder-based receivers~\cite{OShea_PHY_DL_2017,DL_com_over_the_air_2018,Zhu_transceiver_optimization_NN_2019}, 
	DeepWiPHY preserves the well-known baseband signal processing pipeline, which makes it easier to train and more flexible to use without affecting transmitter designs and existing standards/protocols.
	DeepWiPHY uses constellation points as training labels so that 
	there is no need to assume the availability of perfect instantaneous channel state information (CSI)
	during training phase as has been done in~\cite{DeepMIMO2017,DL_LowResolution_MIMO_Klautau_2018,Yang_DL_double_selective_CE_2019,
	AcousticOFDM_DNN_2019,DL_CE_Soltani_2019,ChanEstNet,CNN_Signal_Detection_CongmingFAN2019}.
	This makes DeepWiPHY compatible with real-world data based NN training
	as the perfect CSI could never be obtained.
	
	\textbf{Standard-based synthetic and real-world dataset generation}: 
	We train DeepWiPHY with both synthetic and real-world data. 
	In particular, the synthetic dataset has around 110 million OFDM symbols and
	the real-world dataset has more than 14 million OFDM symbols.
	Our comprehensive synthetic dataset has specifically considered 
	typical indoor channel models and RF impairments. 
	We also collect a real-world dataset under a wide range of signal-to-noise ratio (SNR) levels and at
	various locations in a large office. 
	This real-world dataset is composed of IEEE 802.11ax packets containing random sequences. To create this real-world dataset, we build a passive sniffing-based data collection testbed 
	with a commercial AP, a Samsung mobile phone, and two USRPs. 
	To the best of our knowledge, this is the first work that trains and evaluates 
	a DL-based OFDM receiver with such comprehensive and standard-based
	synthetic/real-world datasets. 
	More importantly, the created datasets and the trained NN models are publicly accessible~\cite{DeepWiPHY_Dataset}, which can provide a useful resource for researchers and all others who work in wireless designs.
	
	\textbf{Comprehensive practical evaluation}: We comprehensively evaluate DeepWiPHY by comparing it to
	a typical reference implementation of the IEEE 802.11ax receiver, under different channel scenarios with 
	both the synthetic and the real-world datasets. 
	In particular, we evaluate DeepWiPHY with different modulation and coding schemes (MCS) such as 
	64 quadrature amplitude modulation (QAM), 256 QAM and 1024 QAM over a wide range of SNR. 
	Most existing work evaluated their proposed DL-based receiver with low spectral efficiency MCSs 
	such as binary phase shift keying (BPSK)~\cite{CNN_Signal_Detection_CongmingFAN2019,DL_80211p_2019}, 
	quadrature phase shift keying or  (QPSK)~\cite{Ye_Data_Driven_DL_CE_SD_OFDM_2018,DL_80211p_2019,OnlineELM_OFDM_Receiver2019,Balevi_One_Bit_DL_2019}.
	Only a few DL-based receivers were tested with higher MCS such as 16 QAM~\cite{Yang_DL_double_selective_CE_2019,RoemNet_16QAM_2019} or 64 QAM~\cite{Gao_Model_Driven_DL_SE_SD_OFDM_2018,
	CE_SignalDetection_Ensemble_DL_2019,Zhu_transceiver_optimization_NN_2019}.
	We also evaluate the proposed DeepWiPHY in terms of both 
	pre-Forward Error Correction (Pre-FEC) bit error rate (BER) and 
	post-Forward Error Correction (Post-FEC) packet error rate (PER), 
	which has not been done in most existing work. Indeed, 
	the performance improvement in terms of PER is the most crucial indicator 
	for any newly proposed PHY layer algorithms~\cite{TgAx_Evaluation_Methodology}.	
	The validation results show that DeepWiPHY achieves comparable performance to the conventional receivers for most use cases and provides considerable SNR gains for certain use cases.

\section{Related work}\label{subsec:related_works}
\textbf{DL-based OFDM receivers}:
OFDM has been an excellent waveform choice for both 4G/5G cellular and Wi-Fi systems, 
and has motivated many DL-based receiver designs. 
In~\cite{Ye_Data_Driven_DL_CE_SD_OFDM_2018}, DL was 
used to approximate the channel estimation and equalization modules.
It showed that the data-driven DNN can provide a more robust channel estimation 
than conventional methods when fewer training pilots are used. 
In~\cite{Gao_Model_Driven_DL_SE_SD_OFDM_2018}, a model-driven
CNN-based receiver was proposed. It particularly leveraged 
the estimate of conventional algorithms as an input. The trained 
CNN demonstrated a superior channel estimation and BER performance
compared to the DNN architecture proposed by~\cite{Ye_Data_Driven_DL_CE_SD_OFDM_2018}. 
Besides, it showed that using the conventional estimate as an input
can help reduce the complexity of the NN architecture of a DL-based receiver. 
Similarly, a single-layer NN was used in~\cite{OnlineELM_OFDM_Receiver2019}
to improve the traditional least squares (LS) channel estimator. 
In particular, its NN-based channel estimator was trained by 
using the conventional LS estimate and taking into account the nonlinearities 
introduced by high-power amplifiers. 
In~\cite{DL_80211p_2019}, an autoencoder was trained to denoise 
the conventional channel estimate in the vehicular communication 
context where IEEE 802.11p was used. However, all these prior works
(except for~\cite{DL_80211p_2019}) do not follow any existing standards and more importantly, all their training/evaluation were only based on limited synthetic data. 
In contrast, our proposed DeepWiPHY not only follows the latest Wi-Fi standard 
but also gets validated by comprehensive synthetic and real-world data.

\textbf{Real-world dataset based DL applications in wireless PHY layer}:
Applying DL in the PHY layer with real data was initiated by designing 
autoencoder-based systems. In~\cite{DL_com_over_the_air_2018}, a two-step learning 
procedure was proposed to train an autoencoder while its over-the-air evaluation 
only showed comparable performance which is 1 dB inferior to a practical conventional baseline.
In~\cite{AutoEncoder_Aoudia_SDR_2019_Jsac}, the authors trained an
autoencoder using an alternating algorithm. Its practical viability was demonstrated 
through hardware implementation but only a few channel realizations 
were used for training and evaluation, which greatly limited its generalization.
Besides autoencoders, some feature-learning-based designs also provide valuable, 
though limited, preliminary over-the-air evaluation results.
In~\cite{Real_data_Underwater_single_carrier_2019}, 
a DL-based receiver was developed for time-varying underwater acoustic channels. 
Its training data was collected during underwater experiments in the South China Sea. 
Nevertheless, the receiver was only trained to work with a single carrier rather than OFDM.
In~\cite{Real_data_CSI_pred_2019}, a combination of CNN and long short-term memory (LSTM)
was exploited to predict CSI under static environments. The dataset used in~\cite{Real_data_CSI_pred_2019} 
however only contains CSI extracted from the medium access control (MAC) layer rather than 
having raw in-phase and quadrature (IQ) samples for PHY layer design as our work does.
Further in~\cite{Real_data_Vehicular_Com_2019}, a real-world dataset
containing IQ samples (only BPSK) was collected in vehicular communication scenarios 
with the IEEE 802.11p standard. An LSTM NN was trained 
to predict the channel quality in terms of SNR while no DL-based channel equalization was explored therein.
Based on the channel estimator trained in~\cite{Gao_Model_Driven_DL_SE_SD_OFDM_2018}, 
a CNN based detection algorithm was further proposed in~\cite{AI_OFDM_CP_Free_Ye_Li_2019} 
for \textbf{cyclic-prefix (CP)-free} OFDM systems. It, however, only showed the BER results of 
two channel propagation scenarios with fixed SNR levels and the conventional minimum mean squared error (MMSE) equalizer 
was not used as a benchmark. 
In contrast to these prior works, our real-world dataset is perfectly standardized and has a 
much larger diversity in both channel scenarios and MCSs.
Moreover, our evaluation involves both pre-FEC BER and post-FEC PER metrics.

\textbf{Other DL-based channel estimation}:
There are also some DL-based channel estimation algorithms that are designed for special scenarios. 
In~\cite{Balevi_One_Bit_DL_2019}, the authors proposed a novel generative supervised DL model for one-bit channel estimation and also designed an autoencoder for one-bit data detection.
In~\cite{Yang_DL_double_selective_CE_2019}, DL was used to 
estimate doubly selective fading channels, where the conventional estimate is used as an input to the developed NN.
In~\cite{AcousticOFDM_DNN_2019}, a DNN-based channel estimator was trained for underwater acoustic OFDM systems. 
In~\cite{CNN_Signal_Detection_CongmingFAN2019}, a CNN-based detector was proposed for a banded linear system using BPSK modulation.
All these works assume that perfect CSI or channel impulse responses are available for training the NN, which is however not true for any real-world based datasets.


\section{Background}
\subsection{IEEE 802.11ax PHY packet format}\label{sec:review_ax}
In this section, we review some key specifications of IEEE 802.11ax PHY packet format. 
Since this work focuses on the single-user single-spatial-stream communication mode, 
we will only present the High-Efficiency Single User (HESU) packet 
which supports a full-band transmission to a single user in 802.11ax.
In particular, we will highlight the key features of the HESU packet, which 
will be used in both the conventional WLAN receiver presented in Sec.~\ref{sec:ax_ofdm_model} 
and the DL-based receiver proposed in Sec.~\ref{sec:DeepWiPHY_architecture}.

\subsubsection{Components of HESU}
\begin{figure}[t!]
	\centering
	\includegraphics[width = 0.48\textwidth]{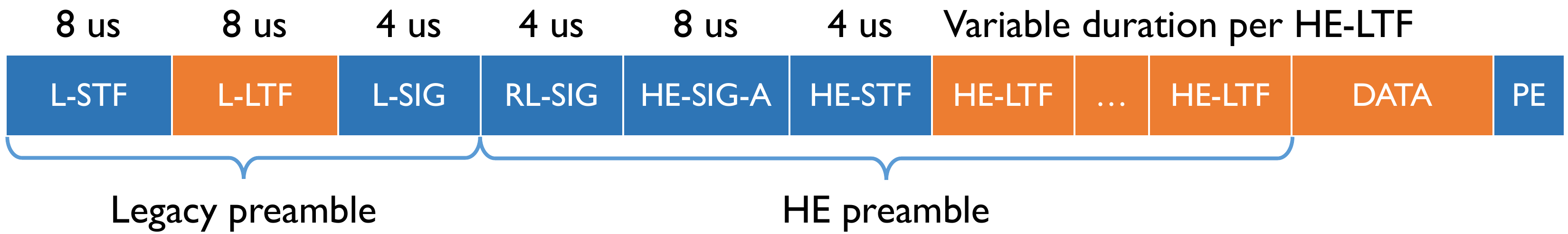}
	\caption{IEEE 802.11ax PHY HESU packet format~\cite{IEEE80211ax}}
	\label{fig:HESU_Frame}
\end{figure}

The HESU packet format is shown in Fig.~\ref{fig:HESU_Frame} and
a list of related acronyms is given in Table~\ref{tab:acronyms}.
As shown in Fig.~\ref{fig:HESU_Frame}, a HESU packet is composed of four parts: (1) legacy preamble, which is included for backward compatibility; (2) HE preamble that supports the new 802.11ax functionality; (3) DATA field that contains FFT-based OFDM symbols; (4) a flexible packet extension field whose content is arbitrary and just provides additional receive processing time.
We will later extract useful raw data from the first three parts as the last one does not provide additional information.

\begin{table}[t!]
	\caption{List of acronyms} 
	\centering
	\label{tab:acronyms}
	\begin{tabular}{ |p{2.0cm}|p{5.cm}|}
		\hline
		L-STF & Legacy Short Training Field\\
		\hline
		L-LTF & Legacy Long Training Field\\
		\hline
		RL-SIG & Repeated Legacy SIGNAL field\\		
		\hline
		HE & High Efficiency \\
		\hline
		HE-SIG-A & HE Legacy SIGNAL field\\					
		\hline
		HE-STF & HE Short Training Field \\
		\hline
		HE-LTF & HE Long Training Field \\
		\hline
		DATA & The Data field \\
		\hline
		PE & Packet Extension field\\
		\hline
		FFT & Fast Fourier transform\\
		\hline
	\end{tabular}\centering
\end{table}

\subsubsection{Feature extraction from HESU}
This work focuses on the receiver design after the time and frequency synchronization blocks.
Thereby, we will only extract useful features for channel estimation, CpeSro correction, and equalization. 
In the legacy preamble, L-LTF is generally used to estimate the variance of 
the noise, L-STF is used for packet detection 
and L-SIG contains transmission parameters of the rest of the packet. Since packet detection 
is not involved in our design, we only extract L-LTF as a feature for later use.
In the HE preamble, RL-SIG and HE-SIG-A contain data pertaining to MAC signaling,
which is not directly relevant to our design. Finally, the DATA field has the OFDM symbols
that are expected to be recovered.
In summary, the useful features for our DL-based design are L-LTF, HE-LTF and DATA fields.
It is worth pointing out that for each OFDM symbol in the DATA field, 
there are pilot tones that are generally used for the CpeSro correction.

\subsubsection{Construction of feature fields of HESU}
We will now elaborate upon constructions of L-LTF, HE-LTF and DATA fields.
L-LTF is composed of a CP followed by two identical training symbols that use 64 FFT. The number of HE-LTF fields is a function of the number of space-time streams. For the single-spatial-stream mode,
there is one HE-LTF per HESU packet.
The default HE-LTF is composed of a CP followed by a training symbol that uses a $N$-Point-FFT,
where $N$ is a function of the bandwidth~\cite{IEEE80211ax}. 
Finally, DATA is composed of $N$-Point-FFT based OFDM symbols. 
As an example, take the case when bandwidth is 20 MHz, for which we have $N=256$. Then for an OFDM symbol 
in the DATA field, there are 234 data tones, 14 null tones, and 8 pilot tones. 
In particular, the pilot sequence of an OFDM symbol is a function of the position of this OFDM symbol relative to the first OFDM symbol in the DATA field. 

\subsection{Conventional OFDM receiver}\label{sec:ax_ofdm_model}
In this section, we review the conventional baseband modules 
that will be replaced by our later proposed DeepWiPHY.
These include channel estimation, CpeSro correction 
and equalization, all in the \textbf{frequency domain}. 

\subsubsection{Signal model}
We denote $\mathsf{x}[k]$ as the symbol carried by the $k$-th subcarrier 
and use subscript T and subscript D to respectively represent the training symbols 
in HE-LTF and the data symbols in the DATA field. Then the received HE-LTF
in the frequency domain can be given as 
\begin{equation}\label{eq:channel_estimation}
\mathsf{y}_\text{T}[k] = \mathsf{h}_\text{T}[k]\mathsf{x}_\text{T}[k] + \mathsf{v}_\text{T}[k],
\end{equation}
where $\mathsf{h}_\text{T}[k]$ is the frequency response at the $k$-th subcarrier and $\mathsf{v}_\text{T}[k]$ is the noise term. Notice that $k$ could refer to either a data or pilot subcarriers (tones)
and $|\mathsf{x}_\text{T}[k]| = 1$, i.e. BPSK symbols.
Similarly, the $m$-th OFDM symbol in the received DATA field can be given as
\begin{equation}\label{eq:CpeSro}
\mathsf{y}_\text{D}[k,m] =\mathsf{h}_\text{D}[k,m]\mathsf{x}_\text{D}[k,m] + \mathsf{v}_\text{D}[k,m],
\end{equation}
where $\mathsf{h}_\text{D}[k,m]$ is the channel frequency response at the $k$-th subcarrier of the $m$-th OFDM symbol in the DATA field and $\mathsf{v}_\text{D}[k,m]$ is the noise term. 

Due to a mismatch of oscillators or Doppler frequency shift, an SRO $\tau$ could exist between receiver and transmitter for symbols in the DATA field, which will contribute to a gradual phase drift in the received signal in the frequency domain~\cite{Chiueh_Baseband_Wireless_MIMO}. 
Moreover, a residual carrier frequency offset (CFO) may also exist for the same reason, which results in
a CPE $\omega[m]$ in the frequency domain, an OFDM symbol dependent phase rotation
to all subcarriers~\cite{Chiueh_Baseband_Wireless_MIMO}. 
As a result, the relationship between $\mathsf{h}_\text{D}[k]$ and $\mathsf{h}_\text{T}[k]$ can be modeled as
\begin{equation}\label{eq:hdht}
\mathsf{h}_\text{D}[k,m] = \mathsf{h}_\text{T}[k]e^{-\mathsf{j}  2\pi\tau\frac{k}{NT_s}}e^{-\mathsf{j}\omega[m]},
\end{equation}
where $\frac{1}{NT_s}$ is the subcarrier spacing. 
In the following, we omit the OFDM symbol index $m$ for simplicity as the following signal processing would be applied to each OFDM symbol in the DATA field separately.

\subsubsection{Channel estimation, pilot tracking, and equalization}
For typical WLAN receivers, channel estimation is 
first performed using HE-LTF to get $\mathsf{h}_\text{T}[k]$. Then CpeSro correction 
is done by jointly estimating CPE and SRO using the 
obtained channel estimates of the pilot subcarriers.

There are many approaches for pilot-based channel estimation, 
among which the LS estimator is the most basic one and has low computational complexity.
In IEEE 802.11ax systems, the LS estimator is given by
\begin{equation}\label{eq:lsCE}
\mathsf{\hat{h}_\text{T}}[k] = \frac{\mathsf{y}_\text{T}[k]\mathsf{x}_\text{T}^*[k]}{\mathsf{x}_\text{T}[k]\mathsf{x}_\text{T}^*[k]} = \mathsf{y}_\text{T}[k]\mathsf{x}_\text{T}^*[k],
\end{equation}
where $*$ is the conjugate operator. 
There are two methods to further improve the quality of the LS channel estimate in (\ref{eq:lsCE}) in the \textbf{frequency domain}. 
The first method is to develop a linear MMSE estimator~\cite{Heath_SP_Book},
which results in high computation cost and requires channel statistics. 
The overhead to obtain these channel statistics can be quite high when the channel is highly dynamic (e.g., the user is fast moving). 
The second method, which is more practical to implement, performs frequency smoothing by using the estimated coherence bandwidth of the channel~\cite{Freq_Smoothing}. Frequency smoothing refers to applying a low-pass filter that spans multiple adjacent subcarriers to the channel estimates $\left\{\mathsf{\hat{h}_\text{T}}[k]\right\}_{k=1}^{K}$ in (\ref{eq:lsCE}). Its achieved gain, however, depends on the channel condition. If adjacent subcarriers are highly correlated, smoothing will result in noise reduction. But for a highly frequency-selective channel, smoothing can degrade the quality of the channel estimate. 
In addition to the above channel estimation methods in the frequency domain, an LS channel estimation in the \textbf{time domain} may provide a better estimate with larger overheads. Please refer to~\cite{Heath_SP_Book} for more details. The performance of frequency smoothing and the time-domain LS channel estimator will be both discussed with the simulation results in Sec.~\ref{sec:evaluation}. 

Pilot symbols exist in both HE-LTF and each OFDM symbol in the DATA field.
By plotting the pilot subcarrier phase differences versus the pilot subcarrier indices, we can obtain the intercept and the slope of its best-fitting line, which corresponds to the CPE and SRO, respectively~\cite{Chiueh_Baseband_Wireless_MIMO}.
Furthermore, the estimated CPE and SRO could be tracked and averaged over multiple OFDM symbols within the same packet. 
Please refer to~\cite{Matlab_CPESTO} for a detailed reference implementation 
of jointly tracking CPE and SRO.
We denote the estimated CpeSro correction coefficient as
$\hat{\mathsf{C}}_\text{CpeSro}[k] = e^{\mathsf{j}2\pi\hat{\tau}\frac{k}{NT_s}}e^{\mathsf{j}\hat{\omega}}$.
Based on (\ref{eq:hdht}), we have the estimated overall channel $\mathsf{\hat{h}_\text{D}}[k]$ given as
\begin{align}\label{eq:overallChannel}
\mathsf{\hat{h}_\text{D}}[k] = \mathsf{\hat{h}_\text{T}}[k]\hat{\mathsf{C}}^*_\text{CpeSro}[k],
\end{align}
which is then substituted into the MMSE equalizer given as 
\begin{align}\label{eq:Coeff_Eq}
\hat{\mathsf{C}}_\text{eq}[k] & = \frac{\mathsf{\hat{h}^*_\text{D}}[k]}{\mathsf{\hat{h}_\text{D}}[k]\mathsf{\hat{h}^*_\text{D}}[k] + \sigma^2},
\end{align}
where $ \sigma^2$ is the variance of the noise, which is usually estimated using L-LTF. 
As a result, an estimation of the transmitted constellation point, denoted as $\hat{\mathsf{x}}_\text{D}[k]$, can be given as
\begin{equation}\label{eq:Prediction}
	\hat{\mathsf{x}}_\text{D}[k] = \hat{\mathsf{C}}_\text{eq}[k] \mathsf{y}_\text{D}[k].
\end{equation}

It is to be noted that there are also many approaches for direct estimation of the equalizer coefficients, as described in \cite{Heath_SP_Book,heath_jr_lozano_2018}. The processing steps presented above constitute a typical one as also adopted in the MATLAB WLAN Toolbox~\cite{Matlab_PER_Example}. 

Having reviewed the baseband signal processing used by a conventional OFDM receiver, in the following section, we will show how a DL-based receiver can be designed to jointly perform the estimation and equalization that are associated with (\ref{eq:lsCE}), (\ref{eq:overallChannel}), (\ref{eq:Coeff_Eq}) and (\ref{eq:Prediction}).

\section{DL-based OFDM receiver}\label{sec:DeepWiPHY_architecture}
In this section, we propose a DL-based aggregated module, named as DeepWiPHY,
to jointly perform the channel estimation, CpeSro correction and equalization, introduced 
in Sec.~\ref{sec:ax_ofdm_model}, all in the \textbf{frequency domain}. 
DeepWiPHY is essentially an aggregated module that follows 
the time and CFO synchronization modules but precedes the 
constellation demapper in a conventional baseband signal processing pipeline.
This makes DeepWiPHY easier to train and more flexible to use. 
We will now delve into the architecture of DeepWiPHY, including the structure of its NN, loss function and input features.

\begin{figure*}[!t]
	\centering
	\includegraphics[width = 0.9\textwidth]{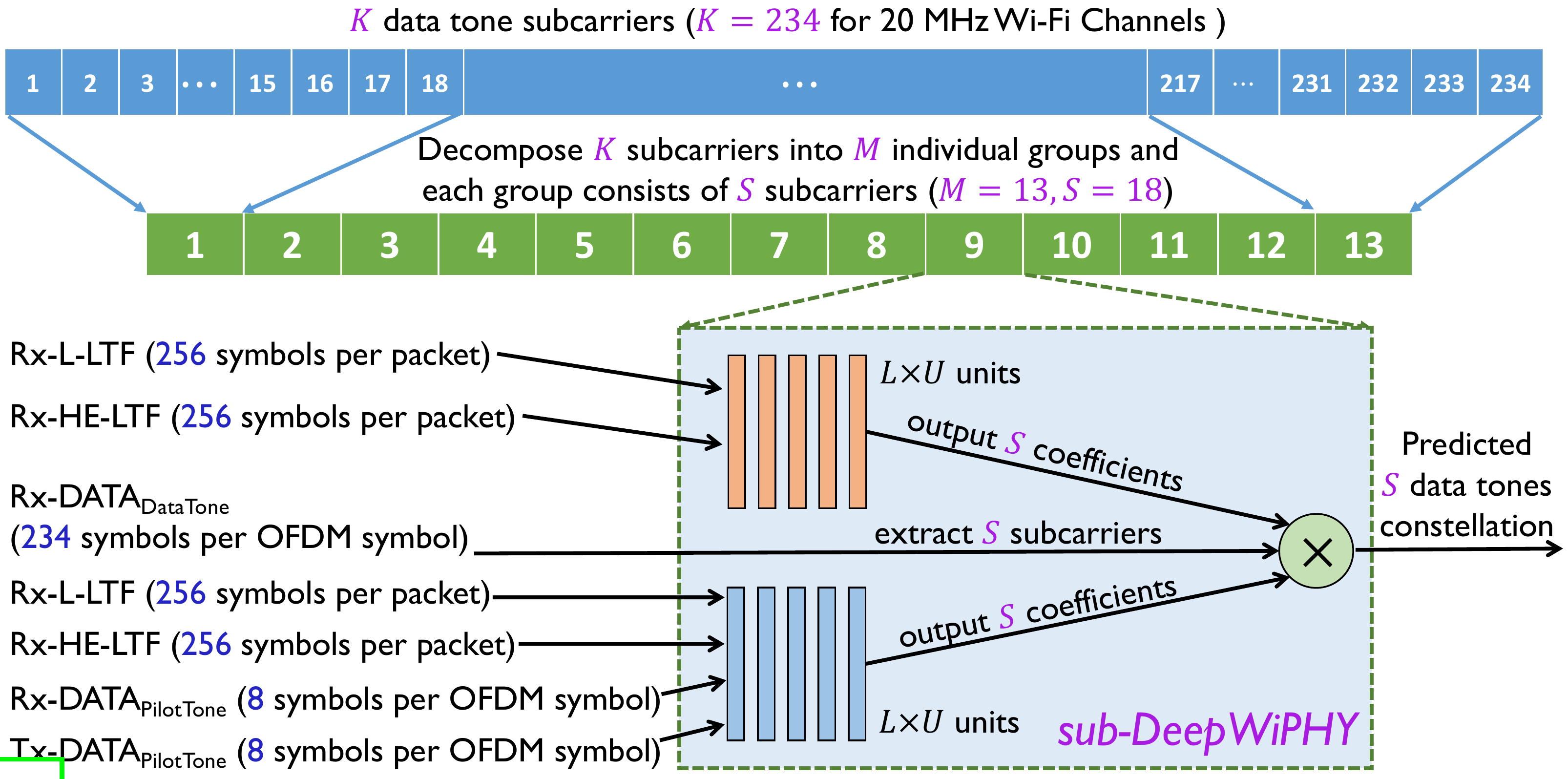}
	\caption{Architecture of DeepWiPHY: there are $M$ sub-DeepWiPHYs and each one performs prediction over $S$ subcarriers such that $K=MS$. Within a sub-DeepWiPHY, the upper DNN mimics an equalizer while the lower DNN mimics a CpeSro correction module.}
	\label{fig:DeepWiPHY_Architecture}
\end{figure*}

\subsection{Parallel NNs construct DeepWiPHY}\label{subsec:DeepWiPHY_architecture_parallel}
In modern wireless communication, OFDM attempts to overcome frequency-selective fading by dividing the entire band into multiple subcarriers and thus the overall wideband signal into multiple narrowband signals. Nevertheless,
the fading in adjacent subcarriers are not completely uncorrelated. To benefit from both the original intention of OFDM 
and the potential correlation among neighboring subcarriers,
we will construct DeepWiPHY with multiple parallel NNs.

We first decompose $K$ data subcarriers into $M$ individual clusters, 
and each cluster consists of $S$ consecutive subcarriers such that $K=MS$,
which is shown in Fig.~\ref{fig:DeepWiPHY_Architecture} for 
an example that has $K=234$, $M=13$ and $S=18$.
After this decomposition, we will use $M$ NNs to separately 
predict the data carried by these $M$ clusters of subcarriers.
We name the NN associated with the $m$-th cluster of subcarriers as 
$m$-th sub-DeepWiPHY. 
We will train the above described $M$ 
sub-DeepWiPHYs separately but with the same dataset 
and use them together to predict all the subcarriers of 
an OFDM symbol (a set of constellation points) in parallel. 
It is worth pointing out that the choice of $M$ and $S$ could be tuned with experiments, which will be studied in Sec.~\ref{sec:evaluation}.  
 
\subsection{Prediction and loss function}
We recall that the conventional estimate of a transmitted QAM symbol (constellation point) requires 
the channel coefficients and the CpeSro correction coefficients, as shown in (\ref{eq:overallChannel}). 
Inspired by this, we use two DNNs in each sub-DeepWiPHY to respectively mimic equalization
and CpeSro correction. We denote the output of the first DNN
as $\left\{\tilde{\mathsf{C}}_\text{eq}[k]\right\}_{k=(m-1)S+1}^{mS}$  and the output of the other
one as $\left\{\tilde{\mathsf{C}}_\text{CpeSro}[k]\right\}_{k=(m-1)S+1}^{mS}$. 
Then we define the DeepWiPHY's prediction of the $k$-th subcarrier's data $\tilde{\mathsf{x}}_\text{D}[k]$ as 
\begin{equation}\label{Prediction_NN}
\tilde{\mathsf{x}}_\text{D}[k] = \tilde{\mathsf{C}}_\text{eq}[k] \tilde{\mathsf{C}}_\text{CpeSro}[k] \mathsf{y}_\text{D}[k].
\end{equation}
The intuition of exploiting the multiplication relationship to design a NN being shaped like (\ref{Prediction_NN}), is to ease the NN learning process by providing it with the natural structure
of the original estimation problem. In~\cite{OShea_PHY_DL_2017}, they formalize this intuition in a concept known as Radio Transformer Networks (RTN) as a way of augmenting DL models with expert propogation domain knowledge. Further discussion on the NN designs is provided in Sec.~\ref{subsec:dicussion_architecture}.
The loss function for the $m$-th sub-DeepWiPHY is defined as the mean squared error (MSE) between the predicted constellation points and their ground truths, which is given as 
\begin{equation}\label{MSE}
L_m = \frac{1}{S}\sum_{k=(m-1)S+1}^{mS} \left\|\tilde{\mathsf{x}}_\text{D}[k] - \mathsf{x}_\text{D}[k]\right\|^2.
\end{equation}
$L_m$ will be backpropagated to update the weights of the two sub-DNNs of the $m$-th sub-DeepWiPHY for $1\leq m\leq M$.

\subsection{NN Architecture of Sub-DeepWiPHY}
All the $M$ sub-DeepWiPHYs share the same input features,
which will be detailed in this subsection. For notational simplicity, 
we use a prefix Rx to denote the received symbols while 
a prefix Tx denotes the transmitted symbols. Besides, we use subscript DataTone
and subscript PilotTone to differentiate the subcarriers of different functionalities.

As reviewed in Sec.~\ref{sec:review_ax},
the key components of a HESU packet that are used for channel estimation, CpeSro correction and equalization
are the L-LTF, HE-LTF, and DATA fields. As depicted in Fig.~\ref{fig:DeepWiPHY_Architecture}, we have one sub-DNN 
mimic an equalizer (including channel estimation) which has 
Rx-L-LTF and Rx-HE-LTF as its inputs. For the second DNN that mimics
the CpeSro correction, it has two extra features, i.e. the transmitted pilot symbols Tx-DATA\textsubscript{PilotTone} and
the received pilot symbols Rx-DATA\textsubscript{PilotTone}. 
Recall that Tx-DATA\textsubscript{PilotTone} is not a constant 
but a function of the position of the associated OFDM symbol in a packet.
Thus, the different Tx-DATA\textsubscript{PilotTone} also encodes information on how far the symbol to be predicted is from HE-LTF where channel estimation is performed.

Furthermore, Rx-DATA, Rx-L-LTF, Rx-HE-LTF are normalized 
such that their Euclidean norm equals the number of symbols contained in that field. Let $\text{dim}(\mathbf{x})$ represent the dimension of a vector $\mathbf{x}$, then its normalized version $\tilde{\mathbf{x}}$ is given as
\begin{equation}
	\tilde{\mathbf{x}}=  \frac{\mathbf{x}}{||\mathbf{x}||}\sqrt{\text{dim}(\mathbf{x})},
\end{equation}
where $\mathbf{x}$ can be Rx-DATA, Rx-L-LTF and Rx-HE-LTF. After this normalization, Rx-DATA\textsubscript{DataTone} and Rx-DATA\textsubscript{PilotTone} are extracted from Rx-DATA.

We have transformed the complex-valued symbols (features) into the real-valued form
by concatenating the real and imaginary parts of the complex-valued
numbers and then use them as the inputs to DeepWiPHY.

Last but not least, each sub-DNN in any sub-DeepWiPHY has $L$ dense layers 
and each layer has $U$ units. Hyperbolic tangent (tanh) is used as the 
activation function for all units because it can produce both positive and negative values, unlike the sigmoid function, and it is a differentiable function having a steeper gradient than the sigmoid function. We do not use dropout because we do not observe any overfitting
according to our evaluation results in Sec.~\ref{sec:evaluation}.

\section{Synthetic dataset generation}\label{sec:synthetic_dataset}
In this section, we summarize the key properties of the synthetic dataset
that is used to train DeepWiPHY. 
This dataset covers typical indoor channel models and nonlinear RF impairments, 
which has not been done in the existing DL-based receiver designs~\cite{Ye_Data_Driven_DL_CE_SD_OFDM_2018,Gao_Model_Driven_DL_SE_SD_OFDM_2018,AI_OFDM_CP_Free_Ye_Li_2019,DL_80211p_2019}.

This synthetic dataset includes 6 indoor spatial channel models that are
specified by 802.11ax (TGax)~\cite{TgAx_Channel_Model}.
We list the delay spreads and cluster parameters of these 
channel models in Table~\ref{tab:ChannelModels}. 
In addition to these channel models, we also consider 5 types of nonlinear RF impairments 
caused by analog filtering, sample clock frequency offset, CFO, 
phase noise, and power amplifier (PA) nonlinearity. All these
nonlinearities are also suggested by 802.11ax (TGax)~\cite{TgAx_Evaluation_Methodology}.
In particular, our synthetic dataset is generated by using three types of combinations of these nonlinearities, which are denoted as Type I, II, III and summarized in Table~\ref{tab:RFimpairment}, in which "\cmark" indicates that the nonlinear impairment is considered in generating simulated data given the type.
More details on the implementations of these channel models and
RF impairments can be found in~\cite{TgAx_Channel_Model,TgAx_Evaluation_Methodology}.


\begin{table}[t!]
	\caption{WLAN indoor channel models~\cite{TgAx_Channel_Model}}
	\label{tab:ChannelModels}
	\centering
	\begin{tabular}{|M{0.8cm}|M{1.4cm}|M{1.4cm}|M{1.5cm}|M{1.5cm}|}
		\rowcolor{gray!20}
		\hline
		Model & Delay spread (ns) & Number of clusters & Taps per cluster & Propagation scenario \\
		\hline
		a & 0 & 1 & 1 &  Flat fading  \\
		\hline
		b & 15 & 2 &  5,7 & Indoor residential \\
		\hline
		c & 30 & 2 & 10,8 & Indoor small office \\
		\hline
		d & 50 & 3 & 16,7,4 & Indoor typical office \\
		\hline
		e & 100 & 4 & 15,12,7,4 & Indoor large office \\
		\hline
		f & 150 & 6 & 15,12,7,3,2,2 & Large space indoor \\
		\hline
	\end{tabular}
\end{table}

\begin{table}[t!]
	\caption{Synthetic RF impairments~\cite{TgAx_Evaluation_Methodology}}
	\label{tab:RFimpairment}
	\centering
	\begin{tabular}{|M{3.7cm}|M{1.cm}|M{1.cm}|M{1.cm}|}
		\hline
		\rowcolor{gray!20}
		Nonlinearity & Type I & Type II & Type III \\
		\hline
		Analog filtering and resampling & \cmark & \cmark & \cmark \\
		\hline
		Sample clock frequency offset & \xmark & \cmark & \cmark \\
		\hline
		Carrier frequency offset & \xmark & \xmark & \cmark \\
		\hline
		Phase noise & \xmark & \cmark & \cmark \\
		\hline
		PA nonlinearity & \xmark & \cmark & \cmark \\
		\hline
	\end{tabular}
\end{table}
	
In summary, there are 18 settings (6 channel models and each has 3 types of RF impairments)
for the synthetic dataset generation. We generate the data under 6 different levels of SNR. 
For each setting, we generate 3200 HESU packets and each packet has 128 symbols in the
DATA field. The composition and relevant parameters of the synthetic dataset are given in Table~\ref{tab:synthetic_dataset}.

\begin{table*}[t!]
	\caption{Summary of synthetic dataset}
	\label{tab:synthetic_dataset}
	\begin{tabular}{|M{1.0cm}|M{1.0cm}|M{1.0cm}|M{1.0cm}|M{1.0cm}|M{1.0cm}|M{1.0cm}|M{1.0cm}|M{1.0cm}|M{3.8cm}|}
		\hline
		\rowcolor{gray!20}
		\multicolumn{10}{|c|}{Data distribution over SNR of synthetic dataset (in million)} \\
		\hline
		QAM & MCS  &  25 dB &  30 dB  &  35 dB &  40 dB &  45 dB &  50 dB &  $\inf$ dB & Total number of OFDM symbols \\
		\hline
		64 & 7 &  7.3728 &  7.3728  &  7.3728 &  7.3728 &  7.3728 &  0 & 9.216  & 46.08 \\
		\hline
		256 & 8 &  0 &  4.608  &  4.608 &  4.608 &  4.608 &  4.608 & 9.216  & 32.256  \\
		\hline
		1024 & 10 &  0 &  4.608  &  4.608 &  4.608 &  4.608 &  4.608 & 9.216  & 32.256 \\
		\hline
		\rowcolor{gray!20}
		\multicolumn{10}{|c|}{Other parameters} \\
		\hline
		\multicolumn{6}{|c|}{Packet format} & \multicolumn{3}{|c|}{Center frequency} & \multicolumn{1}{|c|}{Bandwidth}\\
		\hline
		\multicolumn{6}{|c|}{IEEE 802.11ax HESU format (single-spatial stream)}  & \multicolumn{3}{|c|}{5.25 GHz} &  \multicolumn{1}{|c|}{20 MHz}\\
		\hline
	\end{tabular}\centering
\end{table*}

\section{Real-world data collection}\label{sec:real_world_dataset}
In this section, we present our data collection system.
First, the key hardware components are summarized.
Second, the hardware setup and the methodology of
passive sniffing-based data collection are explained.
Afterward, we elucidate the establishment of a wireless link
that provides the source information for sniffing.
Finally, we detail the procedures for data collection and 
the required post-processing for real-world dataset generation.

\subsection{Key hardware components}

\begin{figure}[t!]
	\centering
	\captionsetup{justification=centering}
	\includegraphics[width = 0.45\textwidth]{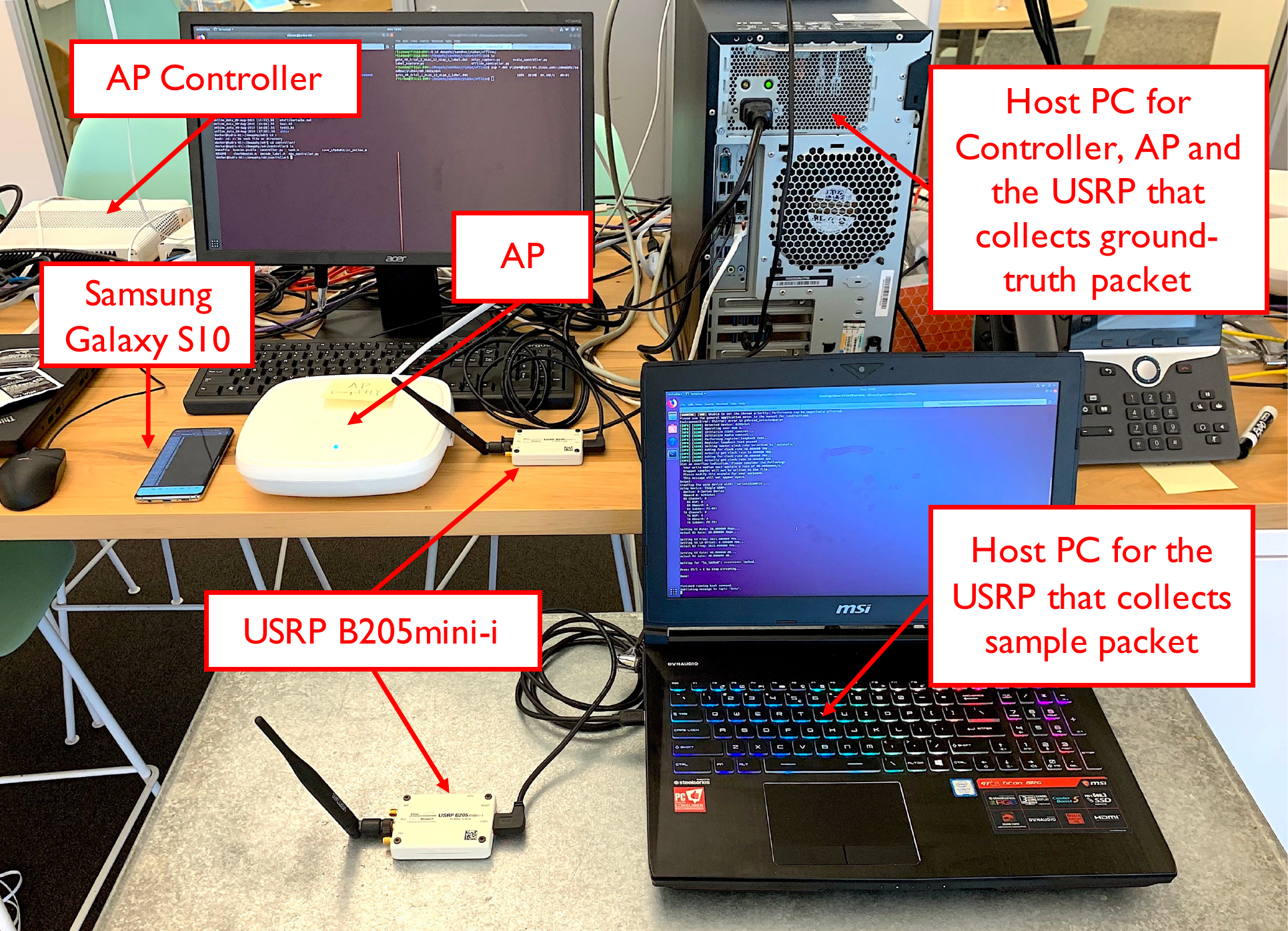}
	\caption{Assembly of key hardware components}
	\label{fig:Assembly}
\end{figure}

As shown in Fig.~\ref{fig:Assembly}, we use one of the latest commercial APs and a Samsung Galaxy S10 mobile phone
to set up a downlink IEEE 802.11ax link. 
In particular, a commercial AP controller is used to configure the number of active antennas of the AP and the MCS used for the established link.
Two B205-mini-i USRPs~\cite{USRPB205} (up to 56 MHz of instantaneous bandwidth)
are used to collect raw IQ samples of packets.
Two host computers that have USB 3.0 compatibility are recommended 
so that the USRP can work properly at the sampling rates required by Wi-Fi channels (at least 20 MHz). 
This setup supports data collection under single-spatial-stream transmitting mode 
for 20/40 MHz Wi-Fi channels. An extension to multiple-spatial-streams mode would require
multiple synchronized USRPs and an extension to 80 MHz Wi-Fi channels would require USRPs that support a higher sampling rate.

\subsection{Methodology of passive sniffing-based data collection}

\begin{figure}[t!]
	\centering
	\captionsetup{justification=centering}
	\includegraphics[width = 0.45\textwidth]{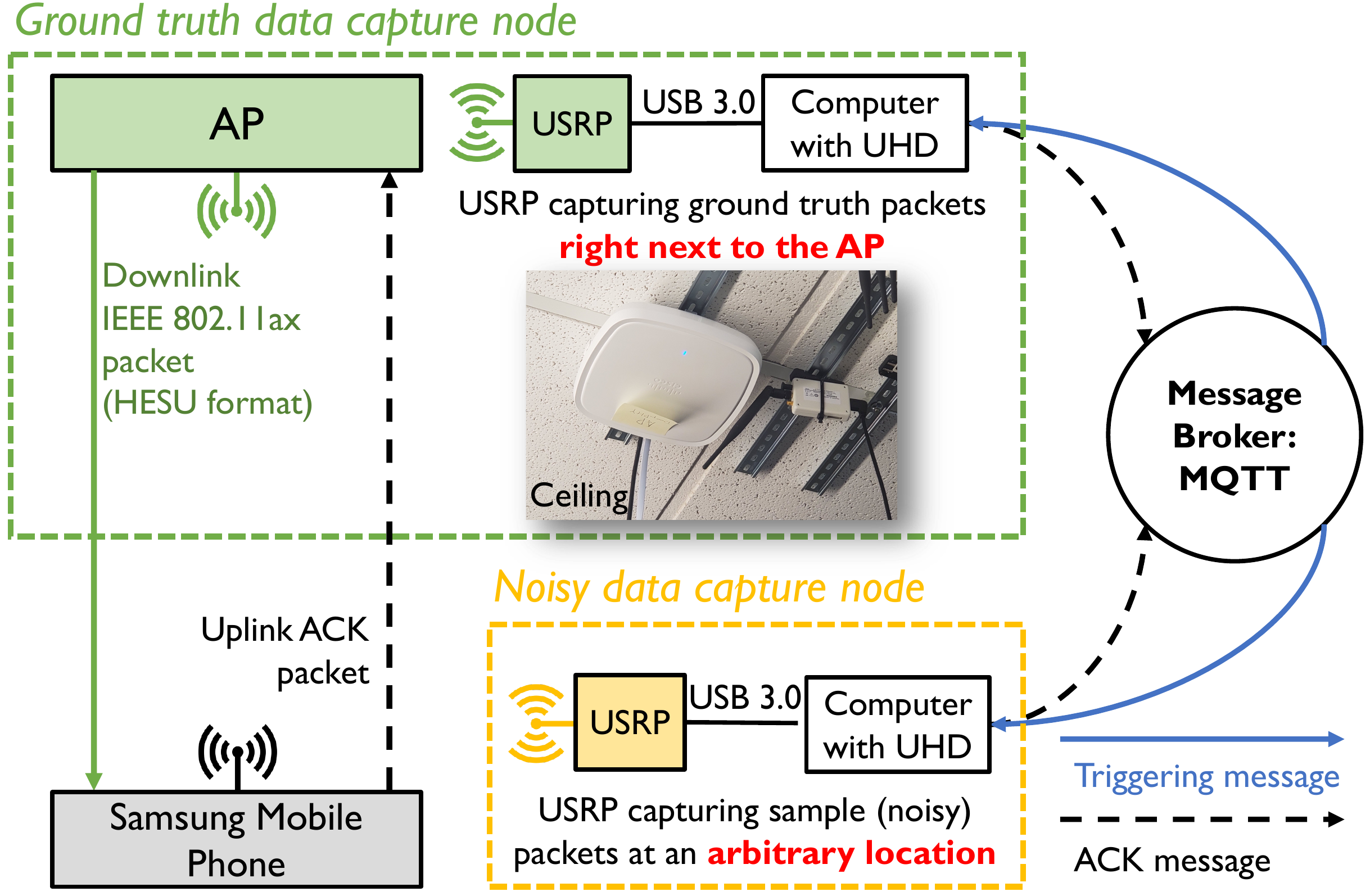}
	\caption{Block diagram of passive sniffing-based data collection}
	\label{fig:Data_Collection_Framework}
\end{figure}

\begin{figure}[t!]
	\centering
	\captionsetup{justification=centering}
	\includegraphics[width = 0.45\textwidth]{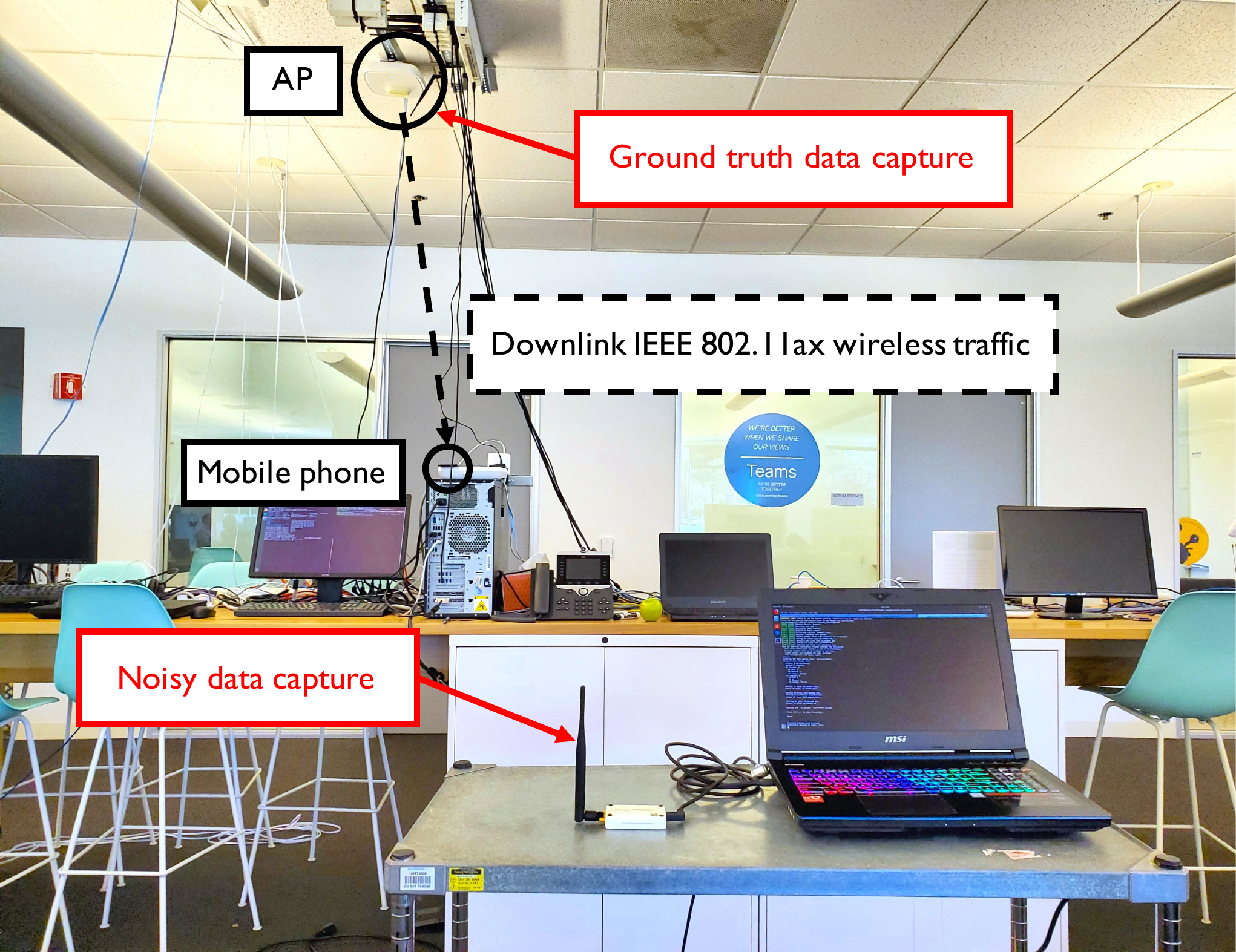}
	\caption{Example of hardware setup for data collection at location 11 in Fig.~\ref{fig:Data_Collection_Map}}
	\label{fig:Loc11}
\end{figure}

A block diagram detailing the hardware connections is given in~Fig.~\ref{fig:Data_Collection_Framework}
and an example setup of the real experiment is shown in~Fig.~\ref{fig:Loc11} for reference.
The key idea of passive sniffing-based data collection is to use two USRPs
to simultaneously (with a tolerable time difference in real implementation) capture the packets
over \textbf{an existing wireless link}. One USRP is used to capture the sample packets while the other is used to capture the 
corresponding ground truth packets.
To be specific, we put one USRP at an arbitrary desired location to capture the sample packets, 
which is referred to as the noisy packets. 
At the same time, we put the other USRP right next to the transmitter (the AP in our testbed), 
as shown in Fig.~\ref{fig:Data_Collection_Framework}, and we refer to its collected packets as 
clean packets.
Such a short distance (around 10 cm) between the AP and the USRP ensures an extremely high SNR (above 50 dB according to the experiments) so that 
the Pre-FEC symbol error rate (SER) is almost zero when we use a conventional WLAN receiver 
to decode those clean packets. As a result, the decoded constellation points can be regarded 
as the corresponding ground truth for the noisy constellation points that are decoded from
the noisy packets.

\begin{figure}[t!]
	\centering
	\includegraphics[width = 0.48\textwidth]{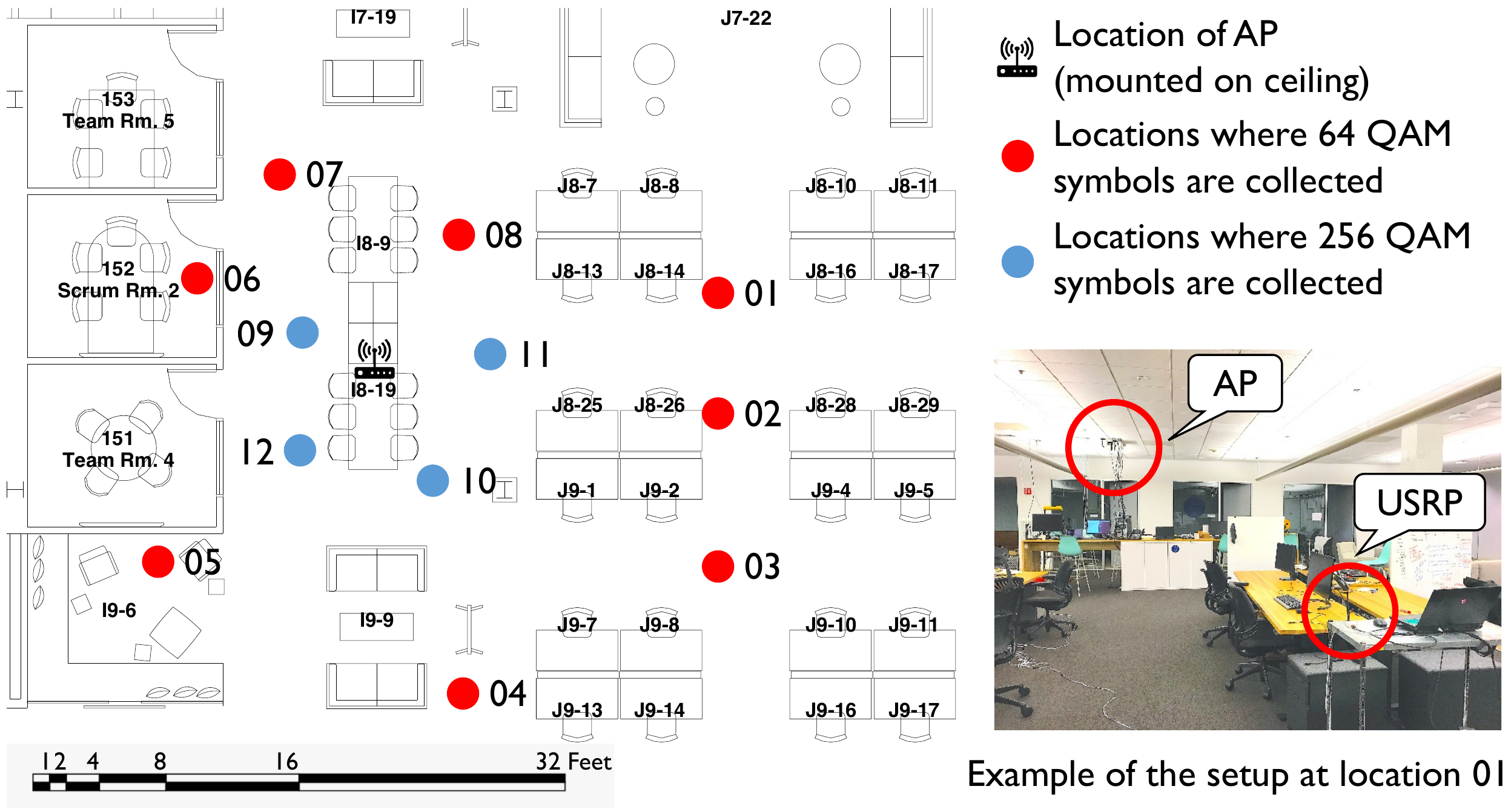}
	\caption{Floor map of the office where the real-world data is collected}
	\label{fig:Data_Collection_Map}
\end{figure}

\subsection{Establishing a wireless link}
In this part, we present our establishment of a wireless link.
We use an AP controller to configure the AP to have a
single active antenna and to transmit a single spatial stream.
The desired MCS index and the center frequency of the used Wi-Fi channel 
are also specified by the controller. 
We perform real-world data collection for MCS 7 (64 QAM) and MCS 8 (256 QAM).
For the higher MCS using 1024 QAM, it is observed in the experiments that 
even with SNR of around 50 dB, a good PER (e.g. below 5\%)
is not always achieved. Therefore, our proposed passive sniffing-based data
collection cannot collect a dataset with accurate labels for the 1024 QAM signal.
A potential solution to this is proposed in Sec.~\ref{sec:online_training} as a future direction.
In this work, we perform the data collection over Wi-Fi channel 156 whose center frequency is 5.825 GHz.
Only one mobile phone is connected to the AP to ensure packets
of HESU format are continually transmitted over the link.
We keep generating sequences of random alphabets and sending
them to the AP via the User Datagram Protocol (UDP) protocol. The AP then packetizes the random sequences 
and sends out the HESU packets over the configured wireless link. 
More importantly, beacon packets are sent by the AP every 102.4 milliseconds (ms). These beacon packets would be used to synchronize and match clean packets and noisy packets later.
The mobile phone will send back an acknowledgment (ACK) message if a packet is correctly received and decoded. 
It is worth pointing out that the mobile phone should be close enough (within 4 meters worked well) to the AP so that the desired MCS could be maintained.

There are two major reasons that we \textbf{combine} off-the-shelf transmitters with 
		software-defined-radio (SDR) receivers to establish the wireless link:
		(1) USRP's allow us to easily obtain the raw IQ samples, while commercial off-the-shelf wireless chipsets rarely provide access to the raw baseband IQ samples
		(2) While it would be easier to get pairs of ground truth packets (from transmitter USRP)
		and their noisy versions (from receiver USRP) under a pure USRP-based framework, leveraging an off-the-shelf transmitter helps to avoid the implementation of non trivial components of a transmitter such as the timing synchronization function (TSF).

\subsection{Procedure for data collection and post processing}

\subsubsection{Quasi-synchronized data collection}
Having established the wireless traffic, we start the data collection 
by broadcasting a triggering message to the USRPs' host computers via 
message queuing telemetry transport (MQTT)~\cite{mqtt}. 
This triggering message is expected to be received by both host computers
at the same time and it contains the required configuration information
for USRP, including center carrier frequency and received gain.
Once the host computer receives the triggering message, it will launch the USRP
and collect raw IQ samples for a period of time, which is set to 5 seconds in our system.
However, due to the different computation ability of host computers and 
the unpredictable network delay, there is an unknown time difference 
between the moments when two USRPs start sniffing data. 
As long as the time difference is less than the
duration of sniffing, there will be common packets collected by both USRPs. 
According to our experiments, this time difference varies from 1 ms to 20 ms, which
is tolerable as our sniffing duration is set to 5 seconds. 
After completing the 5-second data capture, the two host computers will feedback an ACK
message to indicate that the data capture is complete and wait for the next triggering message.
We perform the data collection at 12 locations of a large office shown in Fig.~\ref{fig:Data_Collection_Map}. 
As we can see from Fig.~\ref{fig:Data_Collection_Map},
the locations where 256 QAM symbols are collected are closer to the AP as compared to
those for 64 QAM, because 256 QAM requires a higher SNR.

\subsubsection{Post processing for dataset generation}
Once the raw IQ samples are collected, some post-processing is required
to generate the real-world dataset for training DeepWiPHY later.
As mentioned before, the raw IQ samples collected by different USRPs
are offset by an unknown time difference due to the MQTT network delay and
the different performance of host computers.
To assign ground truth packets to its corresponding noisy ones, 
we perform the following post processing: 
we first decode all the captured clean and noisy packets including 
the beacon packets. By decoding beacon packets, 
we obtain their timestamps which are 8-byte values representing 
the absolute time on the AP when the packet is transmitted. 
We can then approximately calculate the timestamp of each HESU packet
using as reference its closest beacon, the position of the HESU packet with respect to the beacon,
and the sampling rate of the USRPs.
After getting the timestamps for all clean and noisy packets, we can
find the label (ground truth constellation) for each feature vector 
(extracted by decoding the noisy packets) and create a real-world dataset
for future NN training.

The real-world dataset collected is summarized in Table \ref{tab:real_world_dataset}. 
Though technically there is no limitation on the amount of real-world data,
it is non-trivial to capture as much data as in the synthetic methodology. 
\begin{table*}[t!]
	\centering
	\caption{Summary of real-world dataset}
	\label{tab:real_world_dataset}
	\begin{tabular}{ |M{1.0cm}|M{1.0cm}|M{3.5cm}|M{3.5cm}|M{3.5cm}| }
		\hline
		\rowcolor{gray!20}
		QAM & MCS & SNR range  & OFDM symbols per packet & Total OFDM symbols \\
		\hline
		64 & 7 &  27dB - 37dB  & $\approx$ 48 & 9.2 million \\
		\hline
		256 & 8 & 31dB - 39dB & $\approx$ 40 & 4.9 million \\
		\hline
		\rowcolor{gray!20}
		\multicolumn{5}{|c|}{Other parameters} \\
		\hline
		\multicolumn{2}{|c|}{Packet format} & \multicolumn{2}{|c|}{Center frequency} & Bandwidth\\
		\hline
		\multicolumn{2}{|c|}{HESU format} & \multicolumn{2}{|c|}{5.825 GHz}  & 20 MHz\\
		\hline
	\end{tabular}
\end{table*}

\section{Evaluation}\label{sec:evaluation}
In this section, we provide a comprehensive performance evaluation
of our proposed DL-based receiver. We evaluate DeepWiPHY by training and testing it over
20 MHz channels. The benchmarks for DeepWiPHY are the typical
Wi-Fi receivers presented in Sec.~\ref{sec:ax_ofdm_model}.
For evaluation metrics, we use both Pre-FEC BER and PER (Post-FEC).
To evaluate the PER, we use a low-density parity-check (LDPC) decoder that implements the 
min-sum algorithm~\cite{johnsonLDPC} with 20 iterations.
To validate DeepWiPHY under reasonable settings, we design the 
SNR range for testing according to certain industrial analysis reports on
IEEE 802.11ac/ax~\cite{11acSNR_Requirement,Aruba_11ax_WP,Akbilek2019_11ax_Analysis}.
Specifically, the minimum SNR requirements of different MCSs,
for single-spatial-stream transmission over 20 MHz channels, 
are summarized in Table~\ref{tab:SNR_requirement}.
We summarize the specifications of the machines that 
are used for the training in Table~\ref{tab:Spec_Training_Machine}.

\begin{table}[t!]
\caption{Minimum SNR for single-spatial-stream mode over 20 MHz channels~\cite{11acSNR_Requirement,Aruba_11ax_WP,Akbilek2019_11ax_Analysis}}
\label{tab:SNR_requirement}
\begin{tabular}{ |p{1.0cm}|p{1.8cm}|p{1.5cm}|p{2.cm}|}
	\hline
	\rowcolor{gray!20}
	MCS & Modulation & Coding rate & Minimum SNR \\
	\hline
	7 & 64 QAM & 5/6  & $\approx$ 25 dB \\
	\hline
	8 & 256 QAM & 3/4 & $\approx$ 29 dB \\
	\hline
	9 & 256 QAM & 5/6 & $\approx$ 31 dB \\
	\hline
	10 & 1024 QAM & 3/4 & $\approx$ 37 dB \\
	\hline
\end{tabular}\centering
\end{table}
\begin{table}[t!]
			\caption{Specifications of training machine}
			\label{tab:Spec_Training_Machine}
			\begin{tabular}{ |p{1.8cm}|p{5.5cm}|}
				\hline
				GPU model & NVIDIA TESLA P100\\
				\hline
				GPU memory
				& 16 GB \\
				\hline
				CPU model
				& Intel(R) Xeon(R) CPU E5-2637 v4 @ 3.50GHz  \\
				\hline
				CPU core(s) & 8 \\
				\hline
				RAM & 512 GB  \\
				\hline
			\end{tabular}\centering
	\end{table}

\begin{figure*}[!t]
	\begin{subfigure}[t]{0.50\textwidth}
		\centering
		\captionsetup{justification=centering}
		\includegraphics[width = 0.95\textwidth]{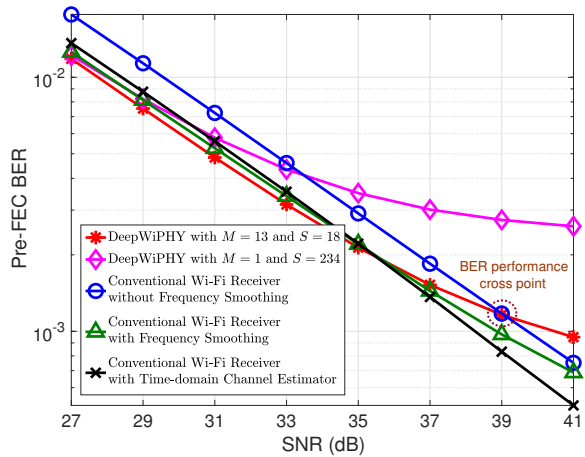}
		\caption{Pre-FEC BER vs SNR with MCS 7, averaged over all channel models.}
		\label{fig:QAM64_MCS7_nSym64_PreFEC_234Tones}
	\end{subfigure}
	\begin{subfigure}[t]{0.50\textwidth}
		\centering
		\captionsetup{justification=centering}
		\includegraphics[width = 0.95\textwidth]{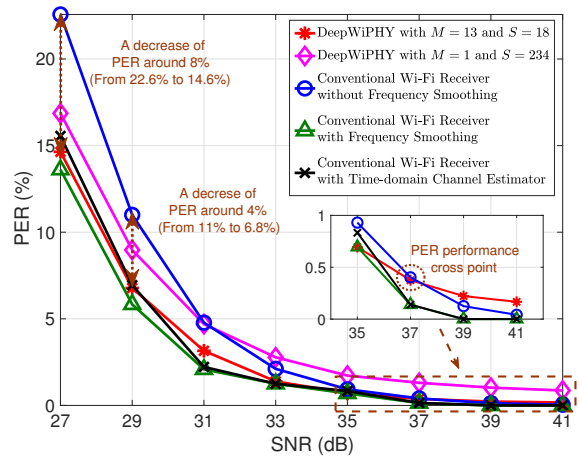}
		\caption{PER vs SNR with MCS 7, averaged over all channel models.}
		\label{fig:QAM64_MCS7_nSym64_PER}
	\end{subfigure}
	\begin{subfigure}[t]{1.0\textwidth}
		\centering
		\captionsetup{justification=centering}
		\includegraphics[width = 1.0\textwidth]{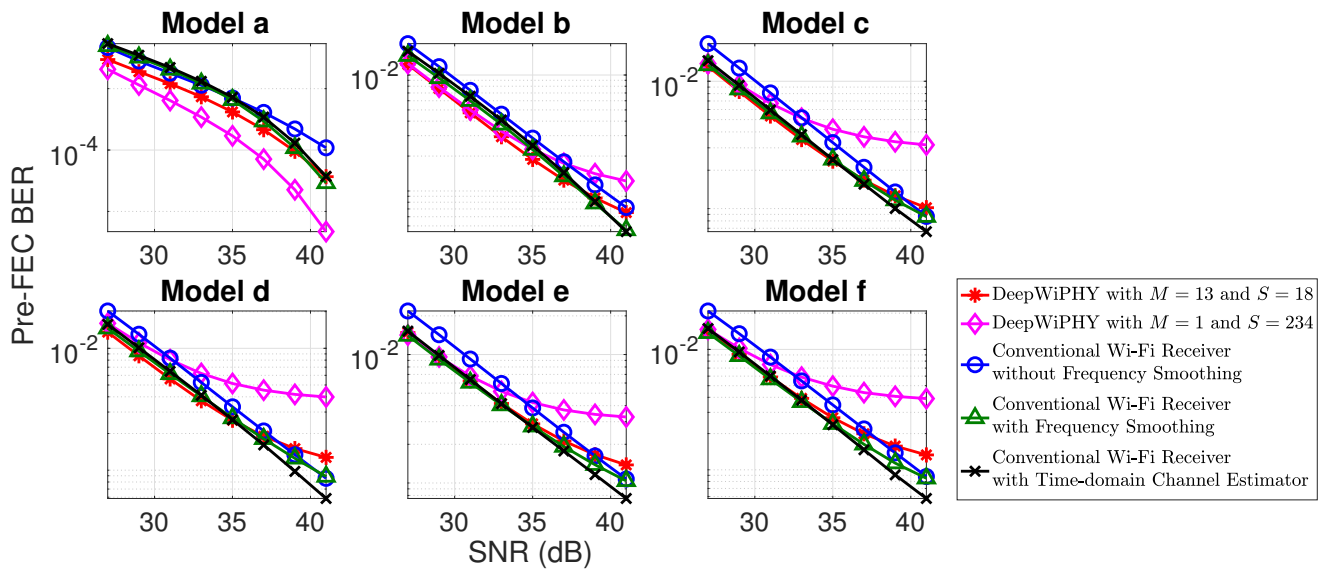}
		\caption{Pre-FEC BER vs SNR with MCS 7 under different channel models}
		\label{fig:QAM64_MCS7_nSym64_PreFEC_Subplot}
	\end{subfigure}
\caption{Synthetic data based evaluation with MCS 7 and DATA field has 64 OFDM symbols. The training parameters are given in Table~\ref{tab:Training_Parameters_Synthetic}. Other evaluation settings are given in Sec.~\ref{subsec:evaluation_synthetic}.}
	\label{fig:QAM64_MCS7_nSym64}
\end{figure*}

\subsection{Discussion on architecture of DeepWiPHY and performance comparisons}\label{subsec:dicussion_architecture}
\subsubsection{Parallel architecture}
as mentioned in Sec.~\ref{sec:DeepWiPHY_architecture}, the motivation to have $M$ parallel sub-DeepWiPHYs is to overcome the frequency-selectivity by dividing the entire band into 
multiple groups and meanwhile to maintain the partial correlation among subcarriers.
The best choice of $M$ could depend on channel condition and would be tuning-based. 
To study this, we compare the average Pre-FEC BER performance of DeepWiPHY using $M=1$ and $M=13$, 
which is given in Fig.~\ref{fig:QAM64_MCS7_nSym64_PreFEC_234Tones}.
Clearly, $M=13$ is significantly better than $M=1$, especially at high SNRs.
Indeed, we also tried $M=234$ but did not observe obvious improvement with respect to $M=13$,
which is shown in Fig.~\ref{fig:architecture_comparision_CNN}. 
These experiments suggest that $M=13$ may be a good choice to
emphasize local correlation and mitigate the frequency smoothing taking place across 
all subcarriers (i.e. $M=1$), and is hence adopted by us for the rest of the evaluation. 
To further investigate this, we show the BER results of each channel
model in Fig.~\ref{fig:QAM64_MCS7_nSym64_PreFEC_Subplot}.
By comparing the BER results of model \textbf{a} (flat fading channel) to those of other models,
we can conclude that DeepWiPHY with $M=1$ is less capable of equalizing multi-path channels, especially in the high SNR region where the multi-path effect becomes more predominant as the delayed paths will not be buried in the noise floor.
In other words, the performance loss of DeepWiPHY with $M=1$ in the high SNR region and under multi-path channels could be interpreted as the application of frequency smoothing beyond the coherence bandwidth of the channel model.

\subsubsection{Performance comparisons}
In Fig.~\ref{fig:QAM64_MCS7_nSym64_PreFEC_234Tones}, 
we also show the Pre-FEC BER performance of conventional receivers with a frequency smoothing filter that spans 9 subcarriers. 
We can see that an appropriate smoothing filter could potentially improve the overall system performance, 
and achieve comparable performance to the receiver with a time-domain LS channel estimator~\cite{Heath_SP_Book}.
It is worth pointing out that the gain achieved by a frequency smoothing filter is not universal and the filter design is often heuristic and empirically determined,
which is a similar shortcoming of any DL-based receivers as the training datasets and NN architectures
are also empirically determined.
In Fig.~\ref{fig:QAM64_MCS7_nSym64_PreFEC_Subplot}, we can see that under channel models \textbf{a} and \textbf{b}, our DeepWiPHY significantly outperforms the conventional receivers (even with frequency smoothing or time-domain channel estimator), and otherwise, it also provides comparable performance.
These observations suggest that we could potentially train different DeepWiPHY models under different channel conditions, which is similar to preparing different smoothing filters for a conventional receiver. This is a common dilemma for DL applications, i.e. the trade-off between generalization and specialization. A potential solution of training a \textit{device-specific} or \textit{location-specific} DL-based receiver is proposed in Sec.~\ref{sec:online_training}.

\subsubsection{Computational complexity and potential enhancement}\label{subsec:exploitCNN}
\begin{figure}
	\centering
	\includegraphics[width = 0.5\textwidth]{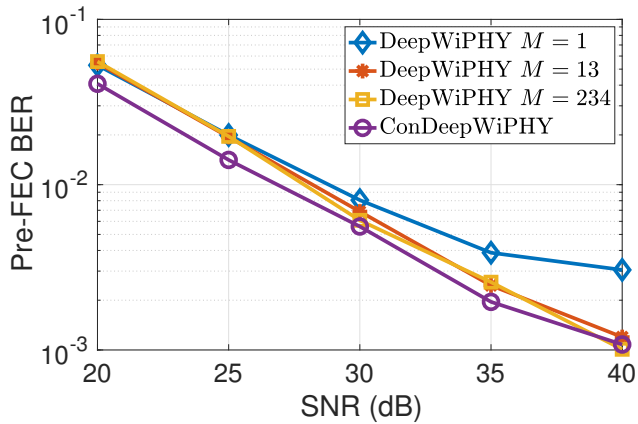}
	\captionsetup{type=figure,justification=centering}
	\caption{Performance comparison between different NN architectures}
	\label{fig:architecture_comparision_CNN}
\end{figure}

To compare the computational cost between DeepWiPHY and conventional receivers,
we provide their execution time (in seconds) for predicting a single OFDM symbol in Table~\ref{tab:complexity_comparison}
This evaluation is performed by MATLAB running in a MacBook Pro with a 3.1 GHz Intel Core i7 CPU and the results are averages of the 2000 executions.
Please note that we utilize the Deep Learning Toolbox in MATLAB which is capable of executing Keras models on GPU, and parallelizing the operation of multiple NNs. According to Table~\ref{tab:complexity_comparison}, we can see that, without using parallel computing, DeepWiPHY with $M=13$ 
is of a quite high computational cost while DeepWiPHY with $M=1$ has a lower cost than all the conventional receivers. This implies that even though the prediction power is enhanced by a larger $M$, the corresponding high complexity may impede its real implementation.
The high complexity of our current architecture comes from the redundancy of having $M$ sub-DeepWiPHYs and each of it composes of two sub-NNs.
One straightforward solution would be exploiting parallel computing 
as the $M$ sub-DeepWiPHYs and their two sub-NNs are fully parallel. Hence the prediction of all 234 tones can be made simultaneously by having all $2M$ NN's perform a forward pass in parallel.
As shown in Table~\ref{tab:complexity_comparison},
the execution time is significantly reduced when using parallel computing.
Another more sophisticated approach is to remove the redundancy by combining those two sub-NNs and further exploiting CNN. The reason for using CNN is because the kernel filter in a convolutional layer naturally lends itself to modeling local correlations in the output, which helps to maintain the correlation between adjacent subcarriers without using parallel sub-DeepWiPHY modules. 
To verify this, we combined the sub-DNNs in DeepWiPHY with $M=1$
and further added convolutional layers at the beginning and the end of it, 
which is termed as ConDeepWiPHY. A sample training result is provided in Fig. \ref{fig:architecture_comparision_CNN},
from which we can see that a better performance is surprisingly achieved
with this much simplified NN architecture.


\begin{table}[t!]
	\captionsetup{type=table,justification=centering}
	\captionof{table}[t]{Execution time for predicting a single OFDM symbol (in seconds)} 
	\label{tab:complexity_comparison}
	\begin{tabular}{|M{3.7cm}|M{1.6cm}|M{1.6cm}|}
		\hline
		\rowcolor{gray!20}
		Types of Receivers & Time without parallel computing & Time with parallel computing \\
		\hline
		DeepWiPHY with $M=1$ & 0.0068 & 0.0032\\
		\hline
		DeepWiPHY with $M=13$ & 0.0805 & 0.0031\\		
		\hline
		Conventional receiver without frequency smoothing & 0.0079 & \diagbox[width=2.03cm, height=0.63cm]{}{}\\
		\hline
		Conventional receiver with frequency smoothing & 0.0081 & \diagbox[width=2.03cm, height=0.63cm]{}{}\\
		\hline
		Conventional receiver with time-domain channel estimator & 0.0094 & \diagbox[width=2.03cm, height=0.63cm]{}{} \\
		\hline
	\end{tabular}\centering
\end{table}

In our evaluation, we do not fine-tune the architecture of DeepWiPHY or the hyperparameter given in Table.~\ref{tab:Training_Parameters_Synthetic} and in Table.~\ref{tab:Training_Parameters_Real_Data}. 
Due to space limitations, please refer to our published dataset for 
the trained models, related codes, and sample training results for ConDeepWiPHY and other NN architectures we ever tested~\cite{DeepWiPHY_Dataset}.
Our evaluation of DeepWiPHY with $M=13$ aims at comprehensively testing the viability of a tailored DL-based receiver that is compliant with a specific standard, and also providing insights into its gains and losses.

\subsection{Evaluation results with synthetic data} \label{subsec:evaluation_synthetic}
\begin{figure*}[t!]
	\begin{subfigure}[t!]{0.50\textwidth}
		\centering
		\captionsetup{justification=centering}
		\includegraphics[width = 0.95\textwidth]{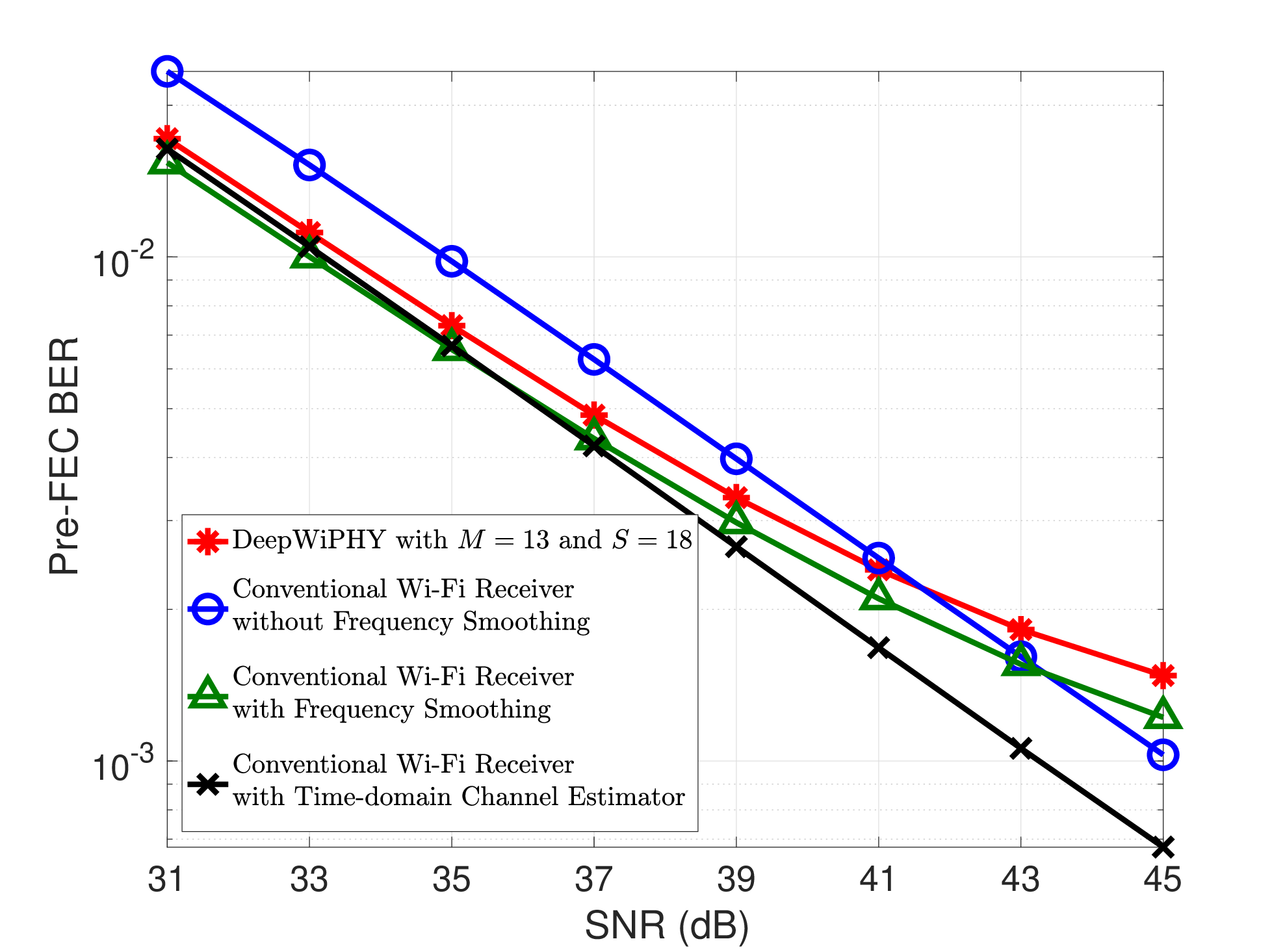}
		\caption{Pre-FEC BER vs SNR with MCS 8}
		\label{fig:QAM256_MCS8_nSym32_PreFEC}	
	\end{subfigure}
	\begin{subfigure}[t!]{0.50\textwidth}
		\centering
		\captionsetup{justification=centering}
		\includegraphics[width = 0.95\textwidth]{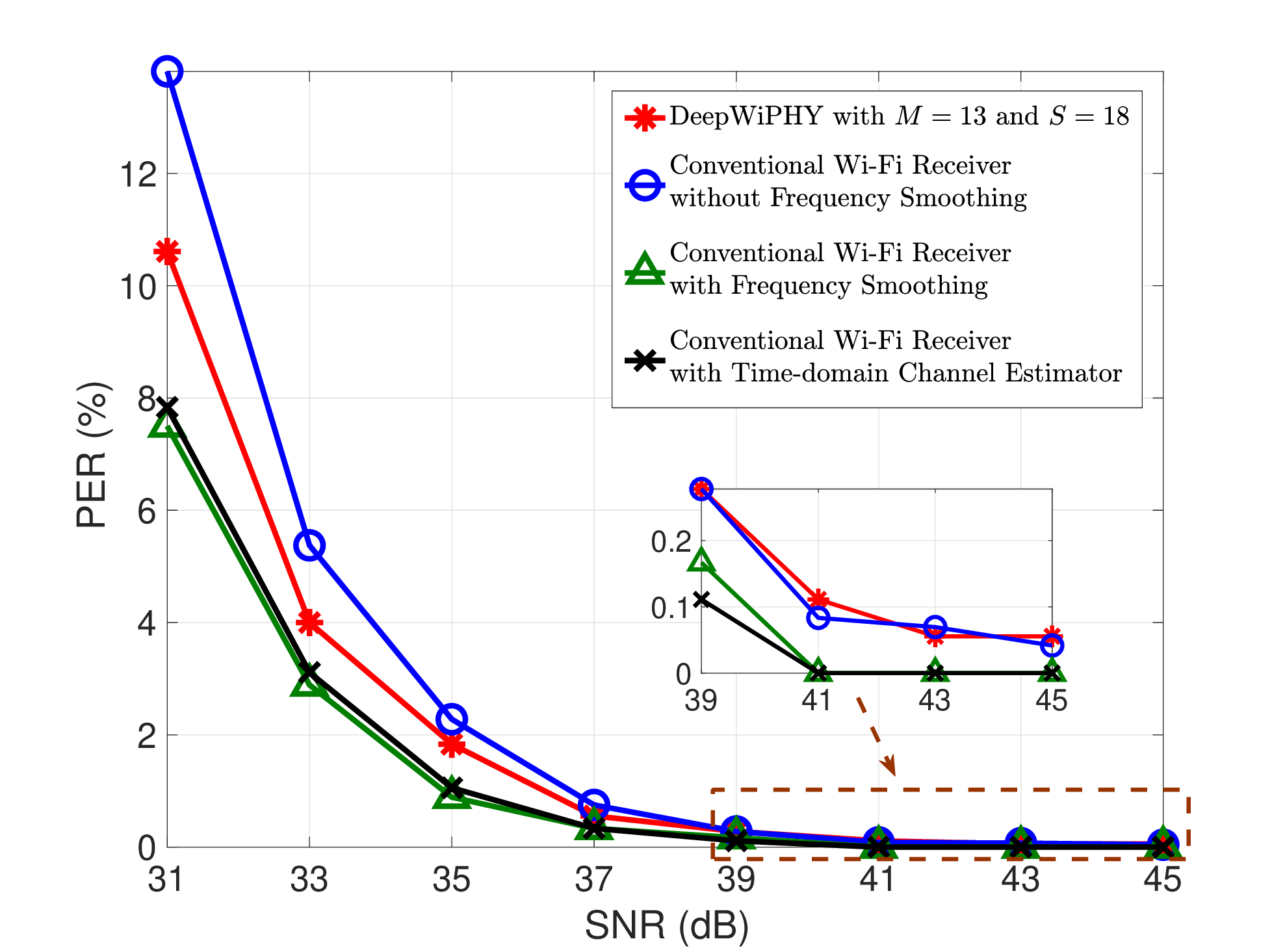}
		\caption{PER vs SNR with MCS 8}
		\label{fig:QAM256_MCS8_nSym32_PER}
	\end{subfigure}
	\caption{Synthetic data based evaluation with MCS 8 and DATA field has 32 OFDM symbols.}
	\label{fig:QAM256_MCS8_nSym32}	
\end{figure*}

\begin{figure*}[t!]
	\begin{subfigure}[t!]{0.50\textwidth}
		\centering
		\captionsetup{justification=centering}
		\includegraphics[width = 0.95\textwidth]{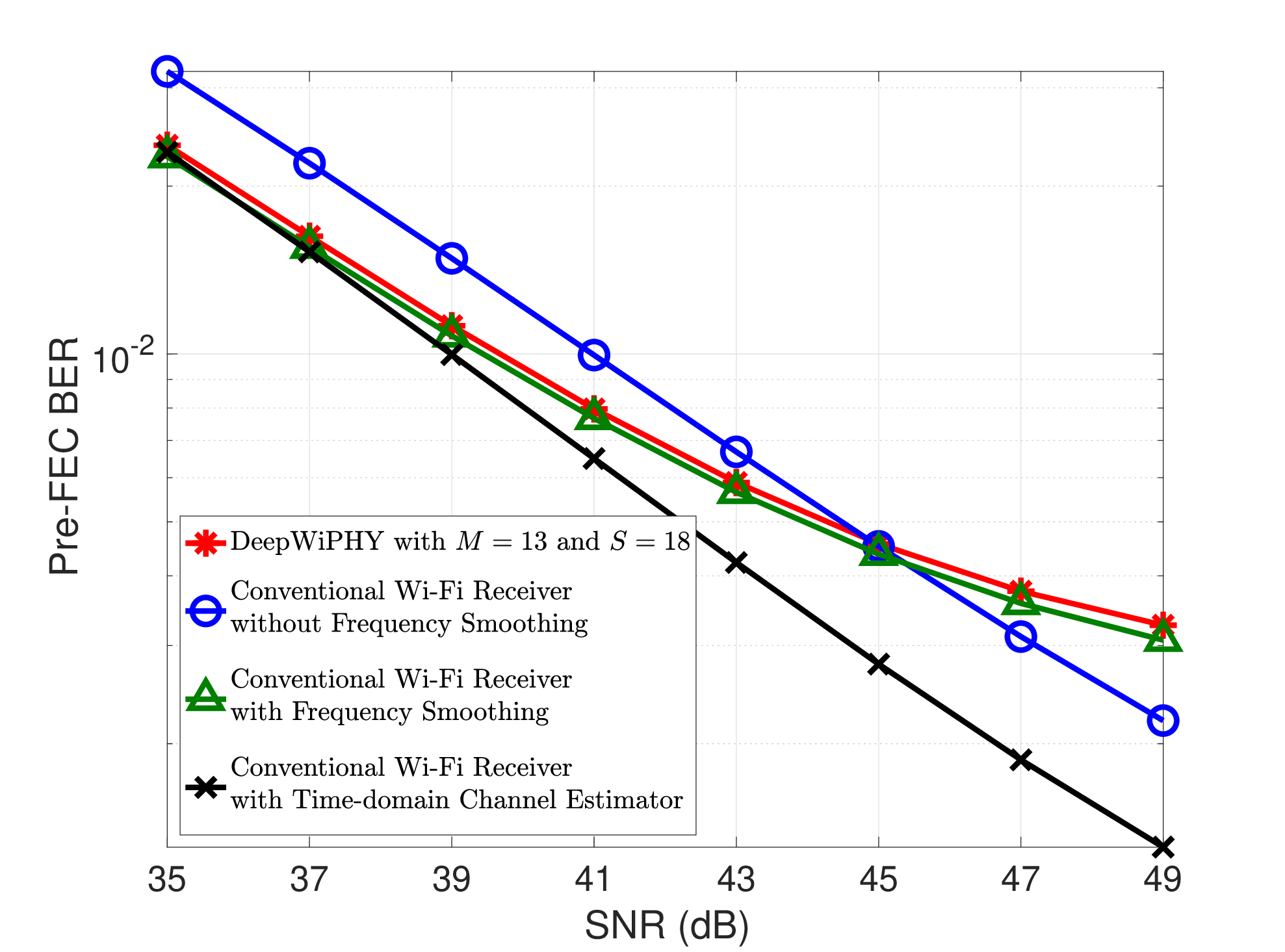}
		\caption{Pre-FEC BER vs SNR with MCS 10}
		\label{fig:QAM1024_MCS10_nSym16_PreFEC}	
	\end{subfigure}
	\begin{subfigure}[t!]{0.50\textwidth}
		\centering
		\captionsetup{justification=centering}
		\includegraphics[width = 0.95\textwidth]{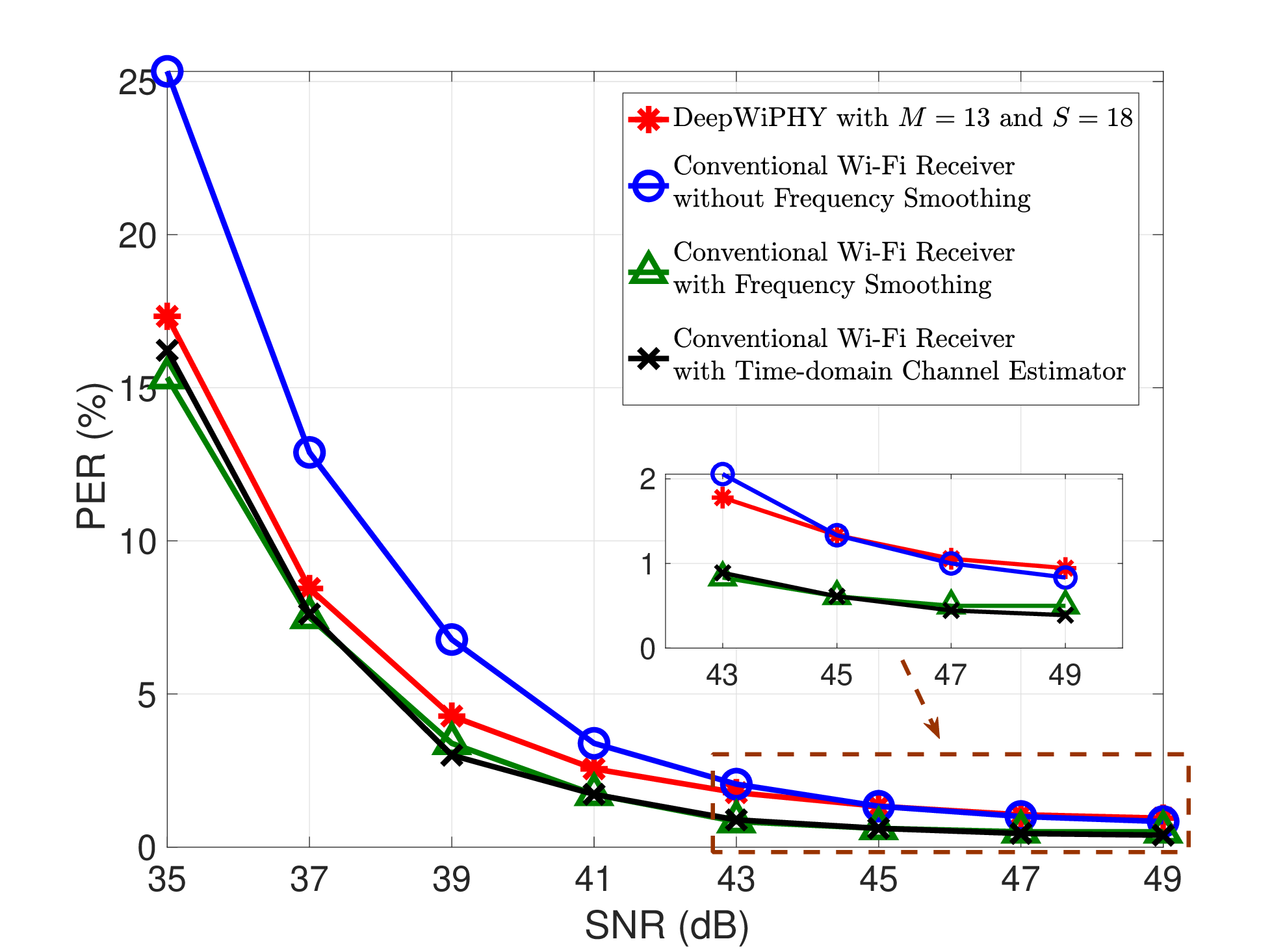}
		\caption{PER vs SNR with MCS 10}
		\label{fig:QAM1024_MCS10_nSym16_PER}
	\end{subfigure}
	\caption{Synthetic data based evaluation with MCS 10 and DATA field has 16 OFDM symbols.}
	\label{fig:QAM1024_MCS10_nSym16}	
\end{figure*}

In this part, we show both the Pre-FEC BER and Post-FEC PER 
performance of DeepWiPHY trained by the synthetic dataset
introduced in Sec.~\ref{sec:synthetic_dataset}. 
The related training parameters are summarized in Table~\ref{tab:Training_Parameters_Synthetic}.
We recall that one epoch refers to the use of the entire dataset to train the NN once, while the iterations is the number of batches
needed to complete one epoch.
Since there is no limitation on the amount of synthetic data we could generate, we use the whole dataset generated in Sec.~\ref{sec:synthetic_dataset} during the training phase
and we regenerate random packets for the testing/validation phase. 
Specifically, the BER or PER result at each 
SNR level in Fig.~\ref{fig:QAM64_MCS7_nSym64_PreFEC_234Tones}, Fig.~\ref{fig:QAM64_MCS7_nSym64_PER}, Fig.~\ref{fig:QAM256_MCS8_nSym32}, and Fig.~\ref{fig:QAM1024_MCS10_nSym16}, 
is averaged over 1800 random packets, which means 100 packets per channel setting
and we have 18 settings in total as presented in Sec.~\ref{sec:synthetic_dataset}.
The numbers of OFDM symbols (in the DATA field) per packet are specified in the captions of figures.
\begin{table*}[t!]
	\caption{Key Training parameters with synthetic data}
	\label{tab:Training_Parameters_Synthetic}
	\begin{tabular}{ |p{7.0cm}|M{2.4cm}|M{2.4cm}|M{2.4cm}|}
		\hline
		Modulation & 64 QAM & 256 QAM & 1024 QAM\\
		\hline
		Training dataset size (number of OFDM symbols)
		& $\approx$ 46 million & $\approx$ 32 million & $\approx$ 32 million\\
		\hline
		Number of sub-DeepWiPHYs: $M$ & \multicolumn{3}{|c|}{13}\\
		\hline
		Number of dense layers in each sub-NN: $L$ & \multicolumn{3}{|c|}{5}\\
		\hline
		Number of units in each dense layer: $U$ & \multicolumn{3}{|c|}{600}\\
		\hline
		Batch size & \multicolumn{3}{|c|}{5120}\\ 
		\hline
		Iterations & 9000 & 6300 & 6300\\
		\hline
		Epochs & 89 & 159 & 317\\
		\hline
		Training time of a sub-DeepWiPHY with $S=18$ & $\approx$ 32 hours & $\approx$ 48 hours & $\approx$ 88 hours\\
		\hline
		Learning rate & \multicolumn{3}{|c|}{1e-5}\\
		\hline
		Optimizer & \multicolumn{3}{|c|}{Adam~\cite{Adam}}\\
		\hline
	\end{tabular}\centering
\end{table*}

In Fig.~\ref{fig:QAM64_MCS7_nSym64}, we show the performance of DeepWiPHY for MCS 7 (64 QAM).
Let us first look at the BER performance shown in Fig.~\ref{fig:QAM64_MCS7_nSym64_PreFEC_234Tones}. 
It can be seen that DeepWiPHY achieves an average SNR gain of 2 dB
over the conventional receiver without frequency smoothing for most SNR levels 
that are suitable for MCS 7 (e.g. below 33 dB). 
Nevertheless, we still notice that when the SNR is high (e.g. beyond 37 dB), 
DeepWiPHY does not provide a lower BER. This observation has been pointed out
in Sec.~\ref{subsec:dicussion_architecture} and it is the motivation that we design DeepWiPHY 
by dividing subcarriers into groups to deal with multi-path channels. 
The observation implies that the current design of DeepWiPHY
is still not fully capable of treating the frequency-selectivity and this is also confirmed in Fig.~\ref{fig:QAM64_MCS7_nSym64_PreFEC_Subplot}, 
where DeepWiPHY only outperforms conventional receivers over the entire SNR range
under channel model \textbf{a} (flat fading channel).
We now focus on the PER performance of DeepWiPHY to further understand its
real achievable gain with respect to the basic benchmark (blue curves). From Fig.~\ref{fig:QAM64_MCS7_nSym64_PER}, several
observations could be highlighted as below: 
(1) At 29 dB, DeepWiPHY brings down PER from 11\% to 6.8\% (around 38\% reduction), which corresponds to an SNR gain of around 1.5 dB. It is worth pointing out that, as a rule of thumb, 10\% PER can be used as a threshold that provides a usable link~\cite{IEEE80211ax}. 
(2) At high SNR levels, e.g. above 37 dB, though DeepWiPHY is not favorable, 
it still achieves comparable performance to the conventional receiver 
with a negligible PER difference (around 0.1\%). 
(3) Moreover, by comparing Fig.~\ref{fig:QAM64_MCS7_nSym64_PreFEC_234Tones} to 
Fig.~\ref{fig:QAM64_MCS7_nSym64_PER}, we can see that the SNR at which 
the two curves cross moves from 39 dB for BER to 37 dB for PER.
This implies that a small BER decrease by DL techniques may not necessarily yield a PER decrease. 
This observation confirms the importance of including PER rather than
only having BER as the metrics for evaluating DL-based receivers.

In Fig.~\ref{fig:QAM256_MCS8_nSym32} and Fig.~\ref{fig:QAM1024_MCS10_nSym16}, 
we provide the performance results (averaged over all channel models) of DeepWiPHY for MCS 8 and MCS 10, respectively.
By vertically comparing the BER and PER performance of DeepWiPHY over these three MCSs, we can find that with the higher modulation order, the smaller performance gain is achieved by DeepWiPHY. This may be due to the fact that a higher modulation order is associated with a more complex classification problem as more constellation points are involved. To deal with higher MCS, more sophisticated NN architecture may be required. These experiments provide an initial study of DL-based receiver using higher MCSs, which is an important omission of the existing work.
Other similar conclusions can be also observed from Fig.~\ref{fig:QAM256_MCS8_nSym32} 
and Fig.~\ref{fig:QAM1024_MCS10_nSym16}, compared to those from Fig.~\ref{fig:QAM64_MCS7_nSym64}.
These confirm the consistency of our analysis and validate our trained DeepWiPHY. Moreover, it is worth pointing out that we train and validate our DeepWiPHY at completely different SNR levels. The consistent performance of DeepWiPHY thus also implies that there is no overfitting in our training.

\begin{table*}[t!]
	\caption{Key training parameters with real-world data}
	\label{tab:Training_Parameters_Real_Data}
	\begin{tabular}{ |p{7.0cm}|M{2.8cm}|M{2.8cm}|}
		\hline
		Modulation & 64 QAM & 256 QAM\\
		\hline
		Training dataset (number of OFDM symbols)
		& $\approx$ 7.9 million & $\approx$ 4.2 million \\
		\hline
		Validation dataset (number of OFDM symbols)
		& $\approx$ 1.3 million & $\approx$ 0.7 million \\
		\hline
		Number of sub-DeepWiPHYs: $M$ & \multicolumn{2}{|c|}{13}\\
		\hline
		Number of dense layers in each sub-NN: $L$ & \multicolumn{2}{|c|}{5}\\
		\hline
		Number of units in each dense layer: $U$ & \multicolumn{2}{|c|}{600}\\
		\hline
		Batch size & \multicolumn{2}{|c|}{5120}\\
		\hline
		Iterations & 1543 & 820 \\
		\hline
		Epochs & 130 & 180 \\
		\hline
		Training time of a sub-DeepWiPHY with $S=18$ & $\approx$ 6 hours & $\approx$ 3 hours \\
		\hline
		Learning rate & \multicolumn{2}{|c|}{1e-6}\\
		\hline
		Optimizer & \multicolumn{2}{|c|}{Adam}\\
		\hline
	\end{tabular}\centering
\end{table*}

\begin{figure*}[t!]
	\begin{subfigure}[t!]{0.50\textwidth}
		\centering
		\includegraphics[width = 0.95\textwidth]{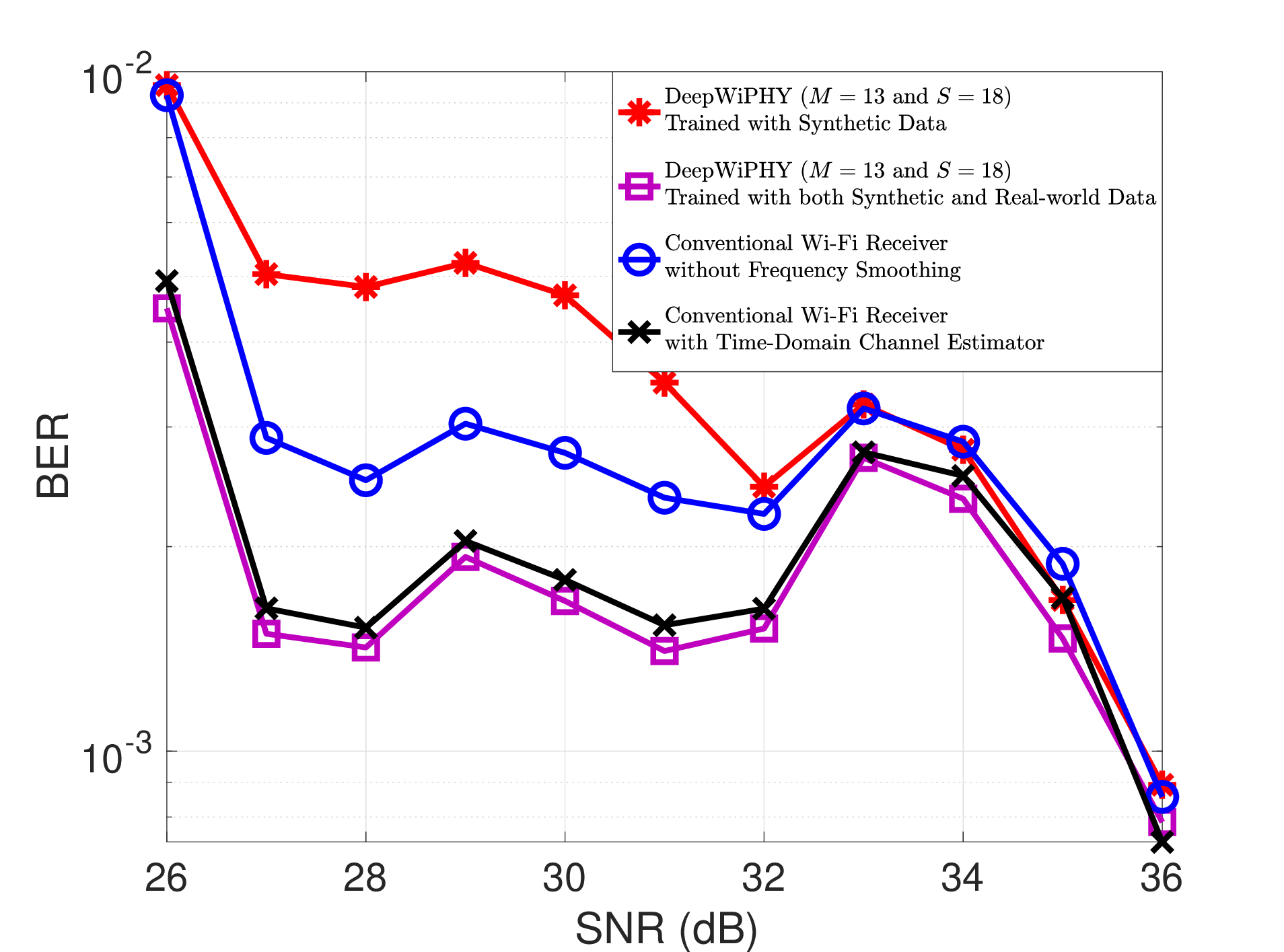}
		\caption{Pre-FEC BER vs SNR with 64 QAM and MCS 7}
		\label{fig:RealData_64QAM_PreFEC}	
	\end{subfigure}
	\begin{subfigure}[t!]{0.50\textwidth}
		\centering
		\includegraphics[width = 0.95\textwidth]{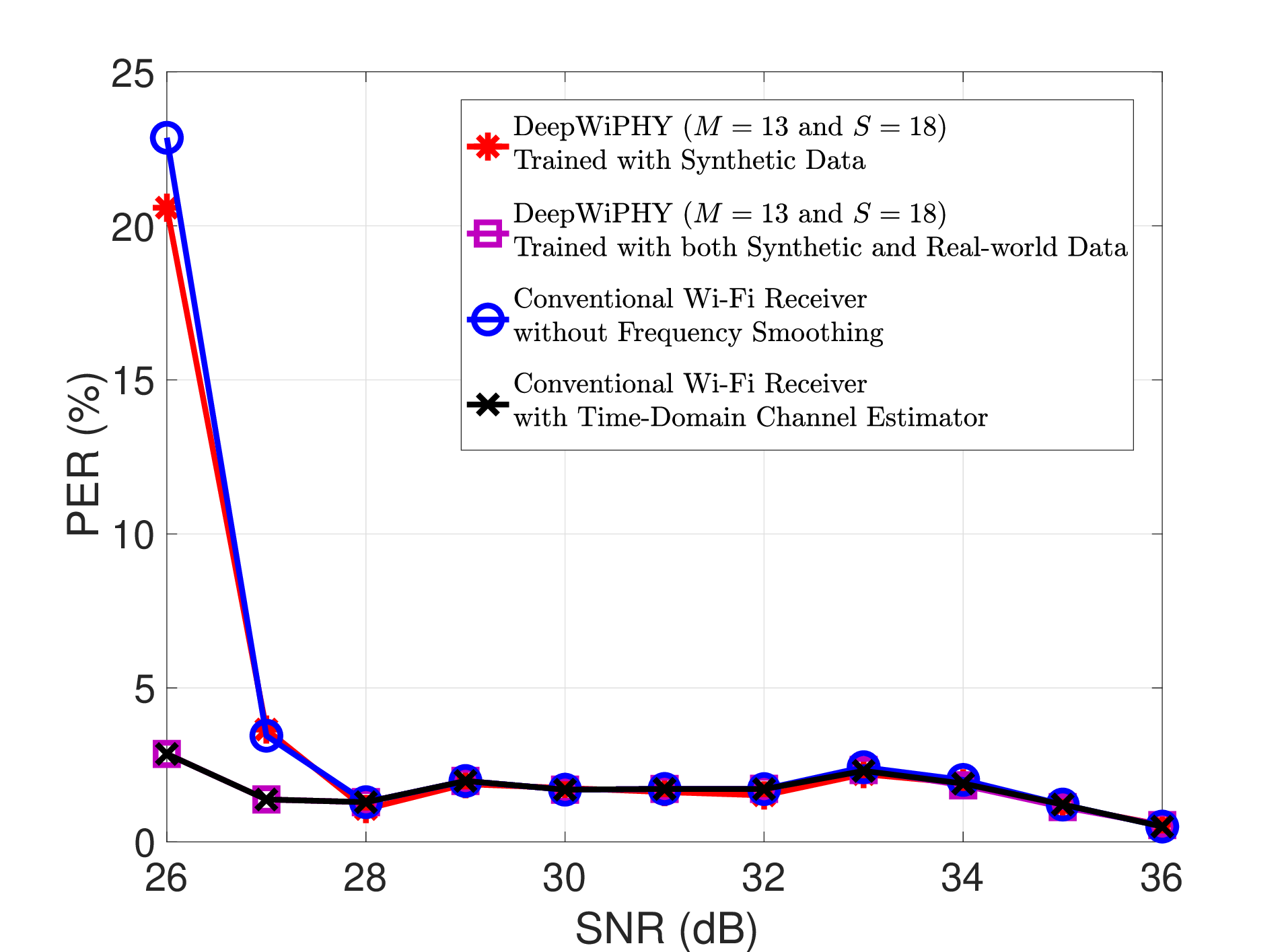}
		\caption{PER vs SNR with 64 QAM and MCS 7}
		\label{fig:RealData_64QAM_PER}
	\end{subfigure}
	\caption{Real-world data based evaluation with MCS 7 and DATA has around 48 OFDM symbols}
	\label{fig:RealData_64QAM}
\end{figure*}

\begin{figure*}[t!]
	\begin{subfigure}[t!]{0.50\textwidth}
		\centering
		\includegraphics[width = 0.95\textwidth]{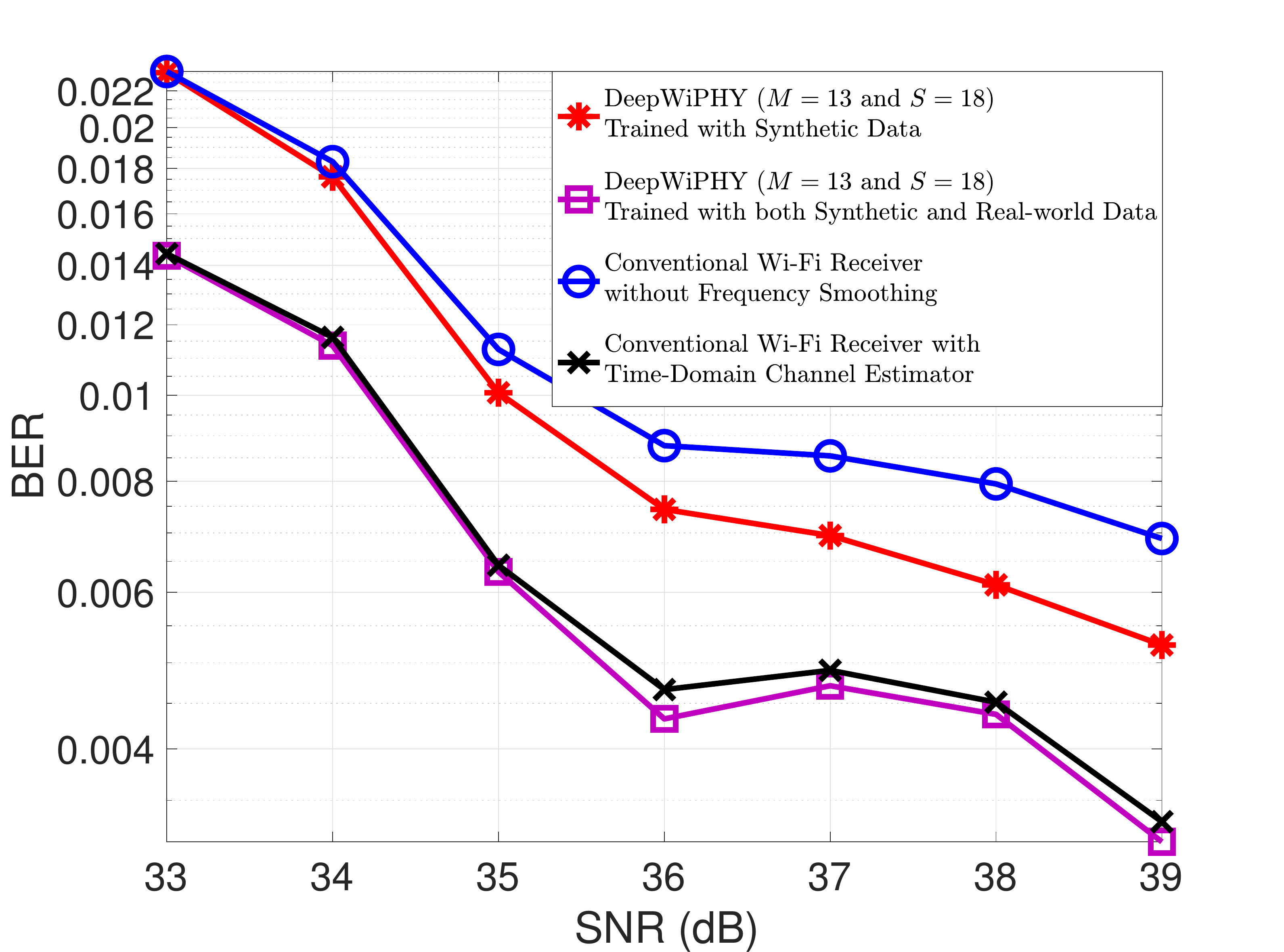}
		\caption{Pre-FEC BER vs SNR with 256 QAM and MCS 8}
		\label{fig:RealData_256QAM_PreFEC}	
	\end{subfigure}
	\begin{subfigure}[t!]{0.50\textwidth}
		\centering
		\includegraphics[width = 0.95\textwidth]{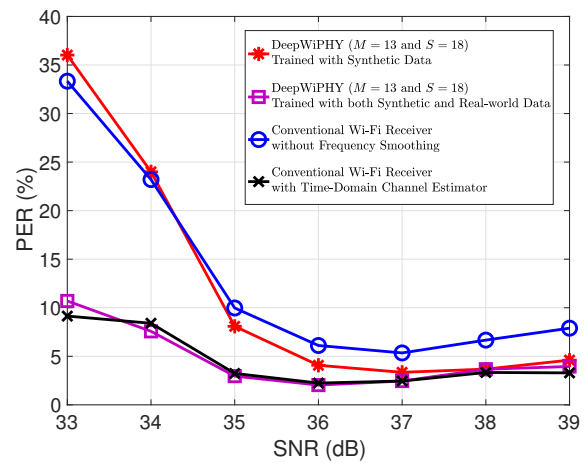}
		\caption{PER vs SNR with 256 QAM and MCS 8}
		\label{fig:RealData_256QAM_PER}
	\end{subfigure}
	\caption{Real-world data based evaluation with MCS 8 and DATA has around 40 OFDM symbols}
	\label{fig:RealData_256QAM}
\end{figure*}

\subsection{Evaluation results with real-world data} \label{subsec:evaluation_real_world_data}
We now present the BER and PER performance
of DeepWiPHY when the real-world data is used for validation. 
Two evaluation methods are used.
The first approach is to directly use the real-world dataset to validate
the DeepWiPHY that is trained only with the synthetic dataset.
The other approach uses both synthetic and real-world datasets during the training phase. 
This means that the DeepWiPHY is first pretrained with the synthetic dataset
and further trained with additional real-world data. 
It is worth pointing out that a \textit{device-specific} or \textit{location-specific} DeepWiPHY module could be trained at each receiver using only real world data, provided it was possible to collect sufficient training data.
The key training parameters are summarized in Table~\ref{tab:Training_Parameters_Real_Data}.
In particular, for the second approach, around 85\% of the captured real-world data
is used during the training phase and the remaining 15\%
is kept for validation. The time-domain channel estimator has similar (or better) performance as the frequency-domain one with frequency smoothing, which can be seen from the previous evaluation.
Accordingly, in the following, we omit the benchmark that the conventional receiver with frequency smoothing for readability.

In Fig.~\ref{fig:RealData_64QAM}, 
we show the performance of DeepWiPHY for real-world 64 QAM symbols. 
In Fig.~\ref{fig:RealData_64QAM_PreFEC}, it can be observed that the DeepWiPHY trained only with synthetic data achieves comparable, 
though slightly worse, performance to the conventional receiver without frequency smoothing. 
On one hand, this suggests that a comprehensive synthetic dataset can train a workable, though not better, receiver. 
On the other hand, this implies the insufficiency of only using a synthetic dataset for training a DL-based receiver.
The observed performance is not unexpected because DL usually achieves better generalization
when the unseen data has the same distribution as the training data. 
It is obvious that the captured real-world dataset does not follow exactly the same distribution as the synthetic dataset.
DeepWiPHY trained by both synthetic and real-world data demonstrates a clear superiority to the conventional receivers in terms of BER. This result well confirms the necessity of using real-world data for training. Using real-world data for training helps to capture the RF impairments that vary from device to device, or some unknown hardware/channel effects that are not commonly simulated. A related online data capturing and training methodology with commercial products is proposed in Sec.~\ref{sec:online_training}.

In Fig.~\ref{fig:RealData_256QAM}, 
we further provide the performance of DeepWiPHY with real-world 256 QAM symbols. 
Similar conclusions can be also observed from Fig.~\ref{fig:RealData_256QAM}, compared to those from Fig.~\ref{fig:RealData_64QAM}.
The demonstrated performance improvement confirms the value of our proposed DL-based receiver.
It is worth pointing out that the curves in Fig.~\ref{fig:RealData_64QAM} and Fig.~\ref{fig:RealData_256QAM} are not smooth because we averaged over all the evaluation results of packets that are captured at different locations as shown in Fig.~\ref{fig:Data_Collection_Map}.
Even though we varied the USRP receiving gains, it is non-trivial to guarantee that the packets could be captured at all SNR levels for every location.

\section{Conclusions \& Potential future directions}\label{sec:discussion}
We have proposed DeepWiPHY, a DL-based OFDM receiver that is
fully compliant with the latest Wi-Fi standard IEEE 802.11ax. 
We have generated a comprehensive synthetic dataset for the NN training 
which has covered typical WLAN indoor channel models and RF impairments. 
In addition to the synthetic data, we have developed a passive sniffing-based 
data collection testbed to capture a large amount of real-world data. 
The value of our proposed DeepWiPHY has been comprehensively validated 
by both synthetic and real-world datasets in comparison with a conventional
Wi-Fi receiver.
Some potential future directions,
which are also applicable to general DL-based OFDM receiver designs, are now discussed as follows.

\subsection{Multiple-spatial-streams mode}
One future direction would be to extend DeepWiPHY for multiple spatial streams.
Indeed, machine learning has already been used for MIMO stream adaptation~\cite{RicHea:Learning-Based-Adaptive-Transmission:14}.
This direction could be extending the channel frequency response $\mathsf{h}_{\text{T}}[k]$ in \eqref{eq:channel_estimation} to be a matrix form and thus convert the multiplication given in
\eqref{eq:Prediction} to matrix multiplications. Moreover, there will be an increase in the size of the input data (recall that the number of HE-LTF fields is a function of the number of spatial streams),
which could increase the size and complexity of the NN architecture of DeepWiPHY. As pointed out in Section~\ref{subsec:exploitCNN}, it could also help by exploiting appropriately designed convolutional filters and/or upsamplers to model the spatial correlation, which is a missing aspect in the current implementation.

\subsection{Online data capturing and training}\label{sec:online_training}
Despite our real-world validation, one could be still concerned that the performance of DeepWiPHY is location or device dependent so that it cannot be trained offline or used by different devices. 
One potential solution is to train DeepWiPHY online at a device locally.
This could be done by having a pretrained NN model loaded into the device and refining this model later. 
By performing cyclic redundancy check (CRC) and validating frame check sequence (FCS), the receiver can identify which part of the received packet is perfectly decoded and accordingly recreate partial transmitted constellation points as labels for training. 
As a result, we could develop a \textit{device-specific} or \textit{location-specific} DL-based receiver 
whose NNs weights change with respect to different hardware and channel conditions. The local online refining of a pre-trained DL-based receiver could provide a non-heuristic but data-driven approach to further improve the performance of Wi-Fi systems.

\bibliographystyle{IEEEtran}
\bibliography{ref}

\begin{IEEEbiography}
	[{\includegraphics[width=1in,height=1.25in,clip,keepaspectratio]{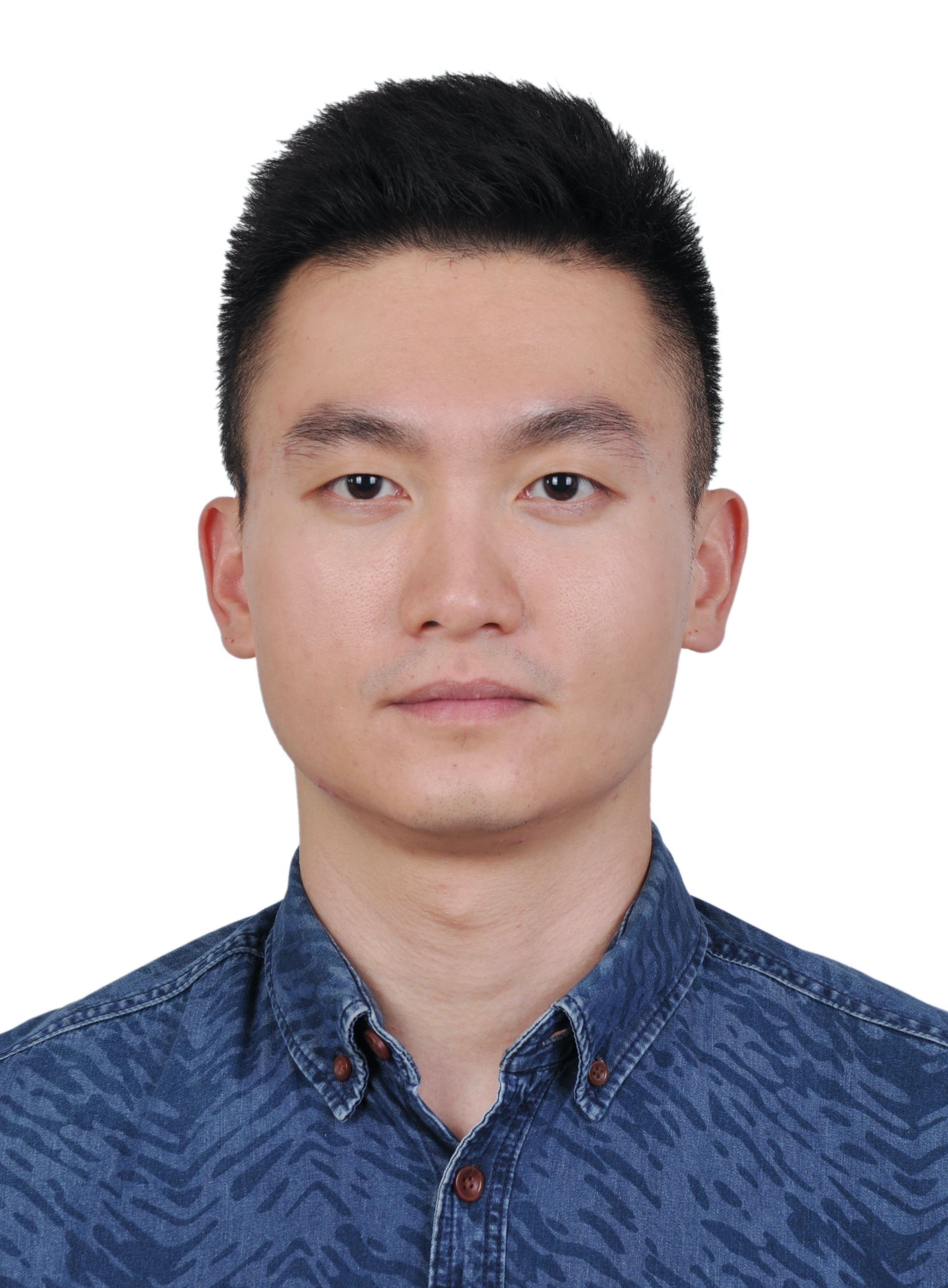}}]
	{Yi Zhang} (S'15) received the B.S. degree from Xi'an Jiaotong University, China in 2014. He received the M.S. degree from Xi'an Jiaotong University, China, and the Engineer's degree from Ecole Centrale de Nantes, France, respectively, in 2017. He is currently pursuing his Ph.D. degree at the University of Texas at Austin, USA. During the summer of 2019, he was interning at the Innovations Lab of Cisco Systems, CA. His research interests include wireless communications and networks, signal processing, reinforcement learning (bandit), deep learning, and wireless system prototyping.
\end{IEEEbiography}

\begin{IEEEbiography}
	[{\includegraphics[width=1in,height=1.25in,clip,keepaspectratio]{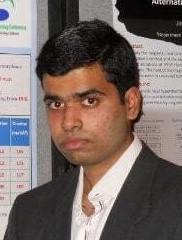}}]
	{Akash Doshi} is currently working towards his PhD at University of Texas at Austin in the Department of Electrical and Computer Engineering. He received his B.Tech (with Honors) in Electrical Engineering from the Indian Institute of Technology (IIT) Bombay in 2018, with a Minor in Computer Science, and his Masters in Electrical and Computer Enginering from the University of Texas at Austin in 2020. His research interests lie mainly in the intersection between wireless networks and machine learning, applying Deep Generative Networks and Reinforcement Learning to various PHY and MAC layer problems for 5G and beyond. 
\end{IEEEbiography}

\begin{IEEEbiography}
	[{\includegraphics[width=1in,height=1.25in,clip,keepaspectratio]{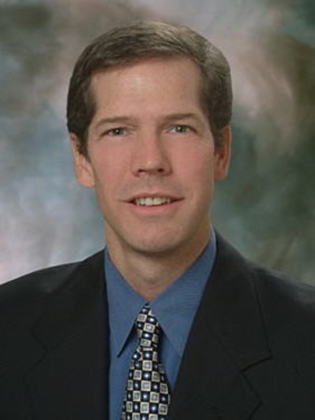}}]
	{Rob Liston} received a BSEE degree from Princeton University and has 25 years experience developing silicon in the graphics, multimedia, and networking domains. He has been with Cisco Systems since 2002, where is a principal engineer in the Intent Based Networking CTO group. He currently works on various aspects of learning and sensing in wireless networking.

\end{IEEEbiography}

\begin{IEEEbiography}
	[{\includegraphics[width=1in,height=1.25in,clip,keepaspectratio]{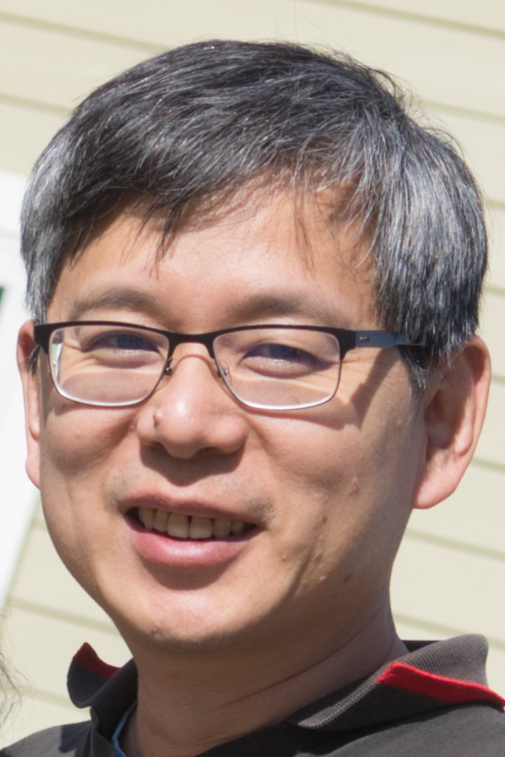}}]
	{Wai-tian Tan} received BS from Brown University, MSEE from Stanford University, and PhD from University of California, Berkeley, all in electrical engineering. He was a researcher at Hewlett Packard Laboratories from 2000 to 2013 working on various aspects of multimedia communications and systems. He has been with Cisco Systems since 2013, where he is a principal engineer in the Innovations Lab within the Intent Based Networking Group. He currently works on various aspects of learning and sensing in wireless networking.
\end{IEEEbiography}

\begin{IEEEbiography}
	[{\includegraphics[width=1in,height=1.25in,clip,keepaspectratio]{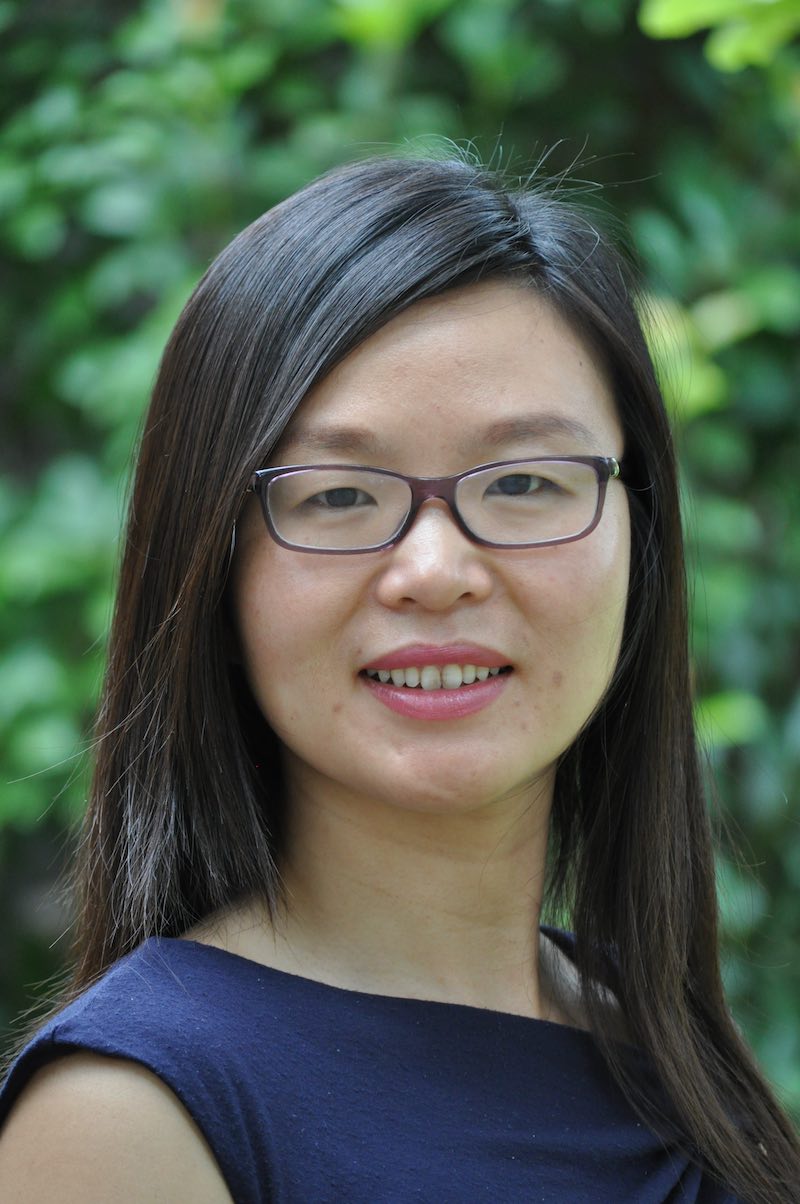}}]
	{Xiaoqing Zhu} is a Sr. Technical Leader at the Innovation Labs of Cisco Systems, Inc. Her research interests include Internet video delivery, real-time interactive multimedia communications, distributed resource optimization, and machine learning for wireless. She holds a B.Eng. in Electronics Engineering from Tsinghua University, Beijing, China. She received both M.S. and Ph.D. degrees in Electrical Engineering from Stanford University, CA, USA. She has published over 90 peer-reviewed journal and conference papers, receiving the Best Student Paper Award at ACM Multimedia in 2007, the Best Presentation Award at IEEE Packet Video Workshop in 2013, and the Best Research Paper Award for VEHCOM 2017. She holds over 35 granted  U.S. patents, with a dozen more pending applications. Xiaoqing has served extensively within the multimedia research community, as TPC member and area chair for conferences, guest editor for special issues of leading journals. and more recently as chair of the MCDIG (Multimedia Content Distribution: Infrastructure and Algorithms) Interest Group in Multimedia Communication Technical Committee (MMTC) and Associate Editor for IEEE Transactions on Multimedia. 
	
\end{IEEEbiography}

\begin{IEEEbiography}
	[{\includegraphics[width=1in,height=1.25in,clip,keepaspectratio]{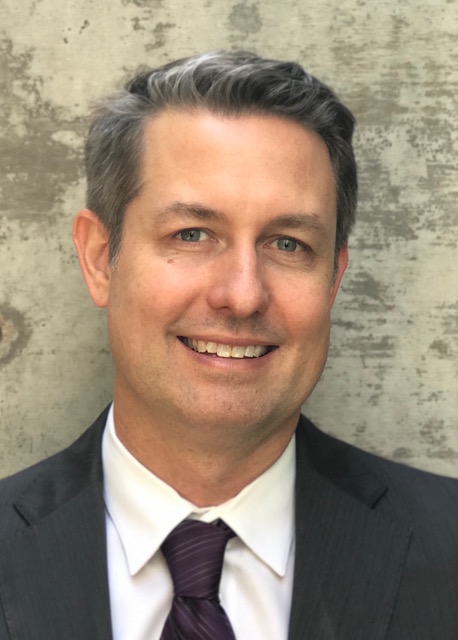}}]
	{Jeffrey G. Andrews} (S'98-M'02-SM'06-F'13) received the B.S. in Engineering with High Distinction from Harvey Mudd College, and the M.S. and Ph.D. in Electrical Engineering from Stanford University.  He is the Cockrell Family Endowed Chair in Engineering at the University of Texas at Austin.  He developed CDMA systems at Qualcomm, and has served as a consultant to Samsung, Nokia, Qualcomm, Apple, Verizon, AT\&T, Intel, Microsoft, Sprint, and NASA.  He is co-author of the books \textit{Fundamentals of WiMAX} (Prentice-Hall, 2007) and \textit{Fundamentals of LTE} (Prentice-Hall, 2010).  He was the Editor-in-Chief of the IEEE Transactions on Wireless Communications from 2014-2016, Chair of the IEEE Communications Society Emerging Technologies Committee from 2018-19, and is the founding Chair of the Steering Committee for the IEEE Journal on Selected Areas in Information Theory, amongst other IEEE leadership roles.
	
	Dr. Andrews is an ISI Highly Cited Researcher and has been co-recipient of 15 paper awards including the 2016 IEEE Communications Society \& Information Theory Society Joint Paper Award, the 2014 IEEE Stephen O. Rice Prize, the 2014 and 2018 IEEE Leonard G. Abraham Prize, the 2011 and 2016 IEEE Heinrich Hertz Prize, and the 2010 IEEE ComSoc Best Tutorial Paper Award.  He received the 2015 Terman Award, the NSF CAREER Award, is an IEEE Fellow, and received the 2019 IEEE Kiyo Tomiyasu technical field award.

\end{IEEEbiography}

\vskip 15\baselineskip 

\begin{IEEEbiography}
	[{\includegraphics[width=1in,height=1.25in,clip,keepaspectratio]{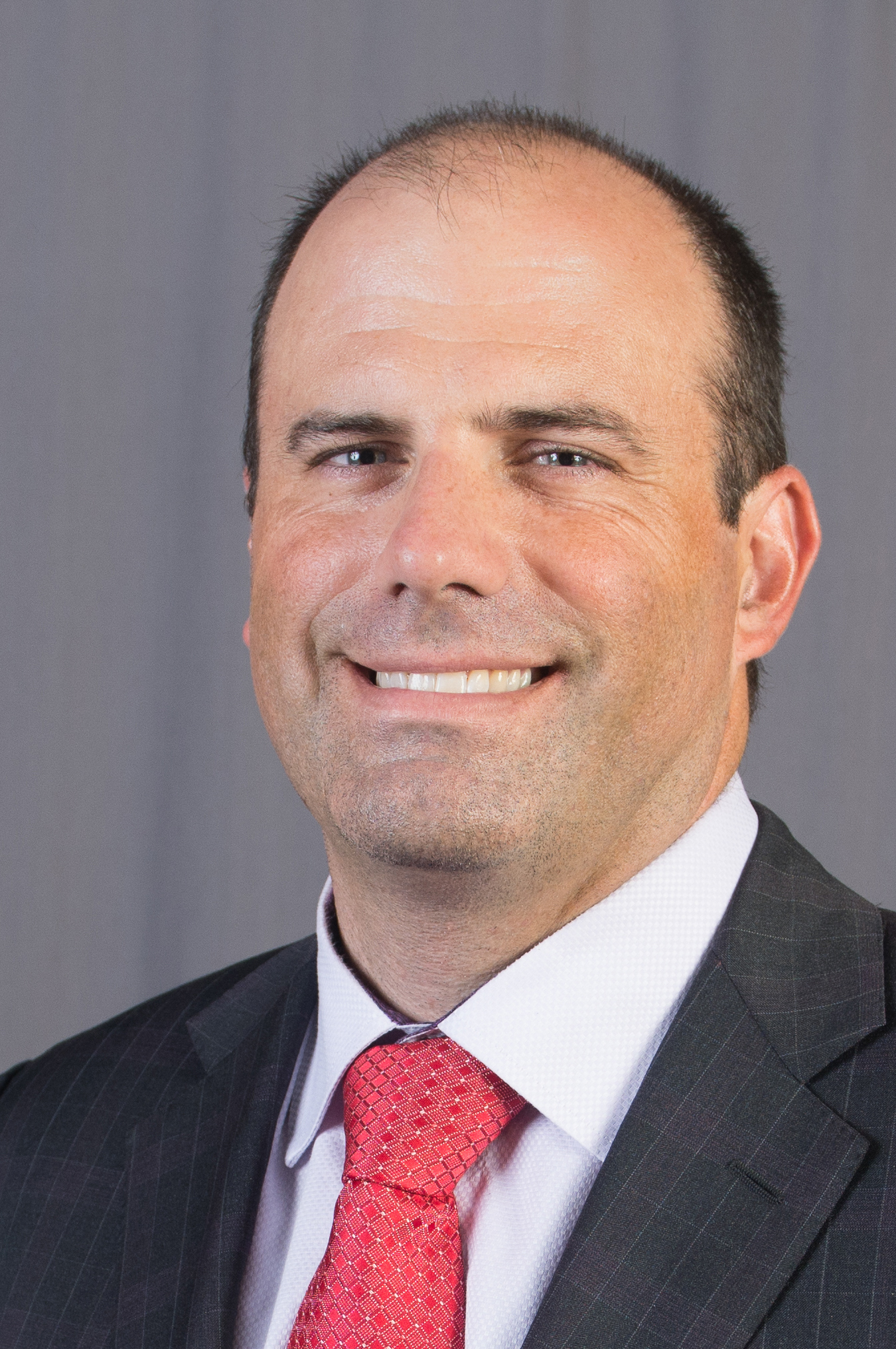}}]
	{Robert W. Heath Jr.} (S'96-M'01-SM'06-F'11)  received the B.S. and M.S. degrees from the University of Virginia, Charlottesville, VA, in 1996 and 1997 respectively, and the Ph.D. from Stanford University, Stanford, CA, in 2002, all in electrical engineering. From 1998 to 2001, he was a Senior Member of the Technical Staff then a Senior Consultant at Iospan Wireless Inc, San Jose, CA where he worked on the design and implementation of the physical and link layers of the first commercial MIMO-OFDM communication system. He is a Distinguished Professor at North Carolina State University. From 2002-2020 he was with The University of Texas at Austin, most recently as Cockrell Family Regents Chair in Engineering and Director of UT SAVES. He is also President and CEO of MIMO Wireless Inc. He authored ``Introduction to Wireless Digital Communication'' (Prentice Hall, 2017) and ``Digital Wireless Communication: Physical Layer Exploration Lab Using the NI USRP'' (National Technology and Science Press, 2012), and co-authored ``Millimeter Wave Wireless Communications'' (Prentice Hall, 2014) and ``Foundations of MIMO Communication'' (Cambridge University Press, 2018). He is currently Editor-in-Chief of IEEE Signal Processing Magazine.
	
	Dr. Heath has been a co-author of a number award winning conference and journal papers including recently the 2016 IEEE Communications Society Fred W. Ellersick Prize, the 2016 IEEE Communications and  Information Theory Societies Joint Paper Award, the 2017 Marconi Prize Paper Award, and the 2019 IEEE Communications Society Stephen O. Rice Prize. He received the 2017 EURASIP Technical Achievement award  and the 2019 IEEE Kiyo Tomiyasu Award. He was a distinguished lecturer and member of the Board of Governors in the IEEE Signal Processing Society. In 2017, he was selected as a Fellow of the National Academy of Inventors. He is also a licensed Amateur Radio Operator, a Private Pilot, a registered Professional Engineer in Texas. 
\end{IEEEbiography}

\end{document}

%% file: input.tex
\usepackage[export]{adjustbox}
\usepackage{textcomp}
\usepackage{algorithm}
\usepackage{algorithmic}
\usepackage{tabularx}
\usepackage[table]{xcolor}
\usepackage{booktabs}
\usepackage{array}
\newcolumntype{P}[1]{>{\centering\arraybackslash}p{#1}}
\newcolumntype{M}[1]{>{\centering\arraybackslash}m{#1}}
\usepackage{multicol}
\usepackage{multirow}
\usepackage{caption}
\usepackage{subcaption}
\usepackage{color}
\usepackage{capt-of}
\usepackage{balance}
\usepackage{graphicx}
\usepackage{epstopdf}
\usepackage{stfloats}
\usepackage[cmintegrals]{newtxmath}
\usepackage{amsfonts}
\usepackage{times}
\usepackage{latexsym}

\usepackage{amssymb}
\usepackage{amsmath}
\usepackage{verbatim}
\usepackage{bm}
\usepackage{cancel}
\usepackage{wrapfig}
\usepackage{listings} 
\usepackage{url}
\usepackage{xcolor}
\usepackage{lettrine}
\usepackage{cite}
\usepackage{pifont}
\usepackage{diagbox}
\newcommand{\cmark}{\ding{51}}%
\newcommand{\xmark}{\ding{55}}%

\renewcommand{\IEEEQED}{\IEEEQEDclosed}
\graphicspath{{figure/}{figures/}}
\IEEEoverridecommandlockouts

%% file: deepPHY_V11.bbl
\begin{thebibliography}{10}
\providecommand{\url}[1]{#1}
\csname url@samestyle\endcsname
\providecommand{\newblock}{\relax}
\providecommand{\bibinfo}[2]{#2}
\providecommand{\BIBentrySTDinterwordspacing}{\spaceskip=0pt\relax}
\providecommand{\BIBentryALTinterwordstretchfactor}{4}
\providecommand{\BIBentryALTinterwordspacing}{\spaceskip=\fontdimen2\font plus
\BIBentryALTinterwordstretchfactor\fontdimen3\font minus
  \fontdimen4\font\relax}
\providecommand{\BIBforeignlanguage}[2]{{%
\expandafter\ifx\csname l@#1\endcsname\relax
\typeout{** WARNING: IEEEtran.bst: No hyphenation pattern has been}%
\typeout{** loaded for the language `#1'. Using the pattern for}%
\typeout{** the default language instead.}%
\else
\language=\csname l@#1\endcsname
\fi
#2}}
\providecommand{\BIBdecl}{\relax}
\BIBdecl

\bibitem{Qin_DL_PHY_Overview_2019}
Z.~Qin, H.~Ye, G.~Y. Li, and B.-H.~F. Juang, ``Deep learning in physical layer
  communications,'' \emph{IEEE Wireless Communications}, vol.~26, no.~2, pp.
  93--99, Apr. 2019.

\bibitem{OShea_PHY_DL_2017}
T.~O'Shea and J.~Hoydis, ``An introduction to deep learning for the physical
  layer,'' \emph{IEEE Transactions on Cognitive Communications and Networking},
  vol.~3, no.~4, pp. 563--575, Dec. 2017.

\bibitem{DL_Modulation_Recognition_GuanGui}
Y.~Wang, M.~Liu, J.~Yang, and G.~Gui, ``Data-driven deep learning for automatic
  modulation recognition in cognitive radios,'' \emph{IEEE Transactions on
  Vehicular Technology}, vol.~68, no.~4, pp. 4074--4077, Apr. 2019.

\bibitem{DeepCodeKim}
H.~Kim, Y.~Jiang, S.~Kannan, S.~Oh, and P.~Viswanath, ``Deepcode: Feedback
  codes via deep learning,'' in \emph{Proceedings of the 32nd International
  Conference on Neural Information Processing Systems}, ser. NIPS'18, 2018, pp.
  9458--9468.

\bibitem{Zhu_transceiver_optimization_NN_2019}
B.~Zhu, J.~Wang, L.~He, and J.~Song, ``Joint transceiver optimization for
  wireless communication {PHY} using neural network,'' \emph{IEEE Journal on
  Selected Areas in Communications}, vol.~7, no.~6, pp. 1364--1373, Jun. 2019.

\bibitem{CNN_End2End_arbitrary_block_length}
N.~{Wu}, X.~{Wang}, B.~{Lin}, and K.~{Zhang}, ``A {CNN}-based end-to-end
  learning framework toward intelligent communication systems,'' \emph{IEEE
  Access}, vol.~7, pp. 110\,197--110\,204, 2019.

\bibitem{DL_com_over_the_air_2018}
S.~D\"orner, S.~Cammerer, J.~Hoydis, and S.~ten Brink, ``Deep learning based
  communication over the air,'' \emph{IEEE Journal of Selected Topics in Signal
  Processing}, vol.~12, no.~1, pp. 132--143, Feb. 2018.

\bibitem{3GPP_V16}
``System architecture for the 5{G} system,'' \emph{document TS 23.501 V16.1.0,
  {3GPP}, Jun. 2019}, pp. 1--219, Jun. 2019.

\bibitem{IEEE80211ax}
``{IEEE} draft standard for information technology -- telecommunications and
  information exchange between systems local and metropolitan area networks --
  specific requirements part 11: Wireless {LAN} medium access control ({MAC})
  and physical layer ({PHY}) specifications amendment enhancements for high
  efficiency {WLAN},'' \emph{{IEEE} P802.11ax/D3.0, June 2018}, pp. 1--682,
  Jul. 2018.

\bibitem{FewThingsOnML}
\BIBentryALTinterwordspacing
P.~Domingos, ``A few useful things to know about machine learning,''
  \emph{Commun. ACM}, vol.~55, no.~10, pp. 78--87, Oct. 2012. [Online].
  Available: \url{http://doi.acm.org/10.1145/2347736.2347755}
\BIBentrySTDinterwordspacing

\bibitem{DeepMIMO2017}
N.~{Samuel}, T.~{Diskin}, and A.~{Wiesel}, ``Deep {MIMO} detection,'' in
  \emph{2017 IEEE 18th International Workshop on Signal Processing Advances in
  Wireless Communications (SPAWC)}, Jul. 2017, pp. 1--5.

\bibitem{DL_LowResolution_MIMO_Klautau_2018}
A.~{Klautau}, N.~{Gonz\'alez-Prelcic}, A.~{Mezghani}, and R.~W. {Heath},
  ``Detection and channel equalization with deep learning for low resolution
  {MIMO} systems,'' in \emph{2018 52nd Asilomar Conference on Signals, Systems,
  and Computers}, Oct. 2018, pp. 1836--1840.

\bibitem{Yang_DL_double_selective_CE_2019}
Y.~Yang, F.~Gao, X.~Ma, and S.~Zhang, ``Deep learning-based channel estimation
  for doubly selective fading channels,'' \emph{IEEE Access}, vol.~7, pp.
  36\,579--36\,589, 2019.

\bibitem{AcousticOFDM_DNN_2019}
R.~{Jiang}, X.~{Wang}, S.~{Cao}, J.~{Zhao}, and X.~{Li}, ``Deep neural networks
  for channel estimation in underwater acoustic {OFDM} systems,'' \emph{IEEE
  Access}, vol.~7, pp. 23\,579--23\,594, 2019.

\bibitem{DL_CE_Soltani_2019}
M.~{Soltani}, V.~{Pourahmadi}, A.~{Mirzaei}, and H.~{Sheikhzadeh}, ``Deep
  learning-based channel estimation,'' \emph{IEEE Communications Letters},
  vol.~23, no.~4, pp. 652--655, April 2019.

\bibitem{ChanEstNet}
Y.~{Liao}, Y.~{Hua}, X.~{Dai}, H.~{Yao}, and X.~{Yang}, ``{C}han{E}st{N}et: A
  deep learning based channel estimation for high-speed scenarios,'' in
  \emph{IEEE International Conference on Communications (ICC)}, May 2019, pp.
  1--6.

\bibitem{CNN_Signal_Detection_CongmingFAN2019}
C.~{Fan}, X.~{Yuan}, and Y.~{Zhang}, ``{CNN}-based signal detection for banded
  linear systems,'' \emph{IEEE Transactions on Wireless Communications},
  vol.~18, no.~9, pp. 4394--4407, Sep. 2019.

\bibitem{DeepWiPHY_Dataset}
\BIBentryALTinterwordspacing
Y.~Zhang, A.~Doshi, R.~Liston, W.~Tan, X.~Zhu, J.~G. Andrews, and R.~W. Heath,
  ``Deep{W}i{PHY}: Synthetic and real-world {IEEE} 802.11ax {OFDM} symbol
  dataset,'' \emph{IEEE Dataport}, 2020. [Online]. Available:
  \url{https://dx.doi.org/10.21227/7h3f-xc81}
\BIBentrySTDinterwordspacing

\bibitem{DL_80211p_2019}
S.~{Han}, Y.~{Oh}, and C.~{Song}, ``A deep learning based channel estimation
  scheme for {IEEE} 802.11p systems,'' in \emph{IEEE International Conference
  on Communications (ICC)}, May 2019, pp. 1--6.

\bibitem{Ye_Data_Driven_DL_CE_SD_OFDM_2018}
H.~Ye, G.~Y. Li, and B.-H.~F. Juang, ``Power of deep learning for channel
  estimation and signal detection in {OFDM} systems,'' \emph{IEEE Wireless
  Communications Letters}, vol.~7, no.~1, pp. 114--117, Feb. 2018.

\bibitem{OnlineELM_OFDM_Receiver2019}
J.~{Liu}, K.~{Mei}, X.~{Zhang}, D.~{Ma}, and J.~{Wei}, ``Online extreme
  learning machine-based channel estimation and equalization for {OFDM}
  systems,'' \emph{IEEE Communications Letters}, vol.~23, no.~7, pp.
  1276--1279, Jul. 2019.

\bibitem{Balevi_One_Bit_DL_2019}
E.~Balevi and J.~G. Andrews, ``One-bit {OFDM} receivers via deep learning,''
  \emph{IEEE Transactions on Communications}, vol.~67, no.~6, pp. 4326--4336,
  Jun. 2019.

\bibitem{RoemNet_16QAM_2019}
H.~{Mao}, H.~{Lu}, Y.~{Lu}, and D.~{Zhu}, ``{R}oem{N}et: Robust meta learning
  based channel estimation in {OFDM} systems,'' in \emph{IEEE International
  Conference on Communications (ICC)}, May 2019, pp. 1--6.

\bibitem{Gao_Model_Driven_DL_SE_SD_OFDM_2018}
X.~Gao, S.~Jin, C.-K. Wen, and G.~Y. Li, ``Com{N}et: Combination of deep
  learning and expert knowledge in {OFDM} receivers,'' \emph{IEEE
  Communications Letters}, vol.~22, no.~12, pp. 2627--2630, Dec 2018.

\bibitem{CE_SignalDetection_Ensemble_DL_2019}
C.~{Ha} and H.~{Song}, ``Signal detection scheme based on adaptive ensemble
  deep learning model,'' \emph{IEEE Access}, vol.~6, pp. 21\,342--21\,349,
  2018.

\bibitem{TgAx_Evaluation_Methodology}
R.~Porat, M.~Fischer, and et~al., ``{IEEE} 802.11ax evaluation methodology,''
  \emph{{IEEE} 802.11-14/0571r11}, pp. 1--47, Nov. 2015.

\bibitem{AutoEncoder_Aoudia_SDR_2019_Jsac}
F.~A. {Aoudia} and J.~{Hoydis}, ``Model-free training of end-to-end
  communication systems,'' \emph{IEEE Journal on Selected Areas in
  Communications}, vol.~37, no.~11, pp. 2503--2516, 2019.

\bibitem{Real_data_Underwater_single_carrier_2019}
Y.~{Zhang}, J.~{Li}, Y.~V. {Zakharov}, J.~{Li}, Y.~{Li}, C.~{Lin}, and X.~{Li},
  ``Deep learning based single carrier communications over time-varying
  underwater acoustic channel,'' \emph{IEEE Access}, vol.~7, pp.
  38\,420--38\,430, 2019.

\bibitem{Real_data_CSI_pred_2019}
C.~{Luo}, J.~{Ji}, Q.~{Wang}, X.~{Chen}, and P.~{Li}, ``Channel state
  information prediction for 5{G} wireless communications: A deep learning
  approach,'' \emph{IEEE Transactions on Network Science and Engineering},
  vol.~7, no.~1, pp. 227--236, 2020.

\bibitem{Real_data_Vehicular_Com_2019}
J.~{Joo}, M.~C. {Park}, D.~S. {Han}, and V.~{Pejovic}, ``Deep learning-based
  channel prediction in realistic vehicular communications,'' \emph{IEEE
  Access}, vol.~7, pp. 27\,846--27\,858, 2019.

\bibitem{AI_OFDM_CP_Free_Ye_Li_2019}
J.~Zhang, C.-K. Wen, S.~Jin, and G.~Y. Li, ``Artificial intelligence-aided
  receiver for a {CP}-free {OFDM} system: Design, simulation, and experimental
  test,'' \emph{IEEE Access}, vol.~7, pp. 58\,901--58\,914, 2019.

\bibitem{Chiueh_Baseband_Wireless_MIMO}
T.-D. Chiueh, P.-Y. Tsai, and I.-W. Lai, \emph{Baseband Receiver Design for
  Wireless MIMO-OFDM Communications}, 2nd~ed.\hskip 1em plus 0.5em minus
  0.4em\relax Wiley-IEEE Press, 2012.

\bibitem{Heath_SP_Book}
R.~W. Heath~Jr., \emph{Introduction to Wireless Digital Communication: A Signal
  Processing Perspective}.\hskip 1em plus 0.5em minus 0.4em\relax Prentice
  Hall, 2017.

\bibitem{Freq_Smoothing}
D.~Katselis, C.~R. Rojas, M.~Bengtsson, and H.~Hjalmarsson, ``Frequency
  smoothing gains in preamble-based channel estimation for multicarrier
  systems,'' \emph{Signal Processing}, vol.~93, no.~9, pp. 2777--2782, 2013.

\bibitem{Matlab_CPESTO}
\BIBentryALTinterwordspacing
{MathWorks, Inc., Natick, MA, USA}. (2018, Oct.) Joint sampling rate and
  carrier frequency offset tracking. [Online]. Available:
  \url{https://www.mathworks.com/help/wlan/examples/joint-sampling-rate-and-carrier-frequency-offset-tracking.html}
\BIBentrySTDinterwordspacing

\bibitem{heath_jr_lozano_2018}
R.~W. Heath~Jr. and A.~Lozano, \emph{Foundations of MIMO Communication}.\hskip
  1em plus 0.5em minus 0.4em\relax Cambridge University Press, 2018.

\bibitem{Matlab_PER_Example}
\BIBentryALTinterwordspacing
{MathWorks,~Inc., Natick, MA, USA}. (2019, Oct.) 802.11ax packet error rate
  simulation for single user format. [Online]. Available:
  \url{https://www.mathworks.com/help/wlan/examples/802-11ax-packet-error-rate-simulation-for-single-user-format.html}
\BIBentrySTDinterwordspacing

\bibitem{TgAx_Channel_Model}
J.~Liu, R.~Porat, and et~al., ``{IEEE} 802.11ax channel model document,''
  \emph{{IEEE} 802.11-14/0882r4}, pp. 1--10, Sept. 2014.

\bibitem{USRPB205}
\BIBentryALTinterwordspacing
{Ettus Research, Santa Clara, CA, USA}. (2019, Jan.) {USRP} {B}205-mini-i.
  [Online]. Available:
  \url{https://www.ettus.com/all-products/usrp-b205mini-i/}
\BIBentrySTDinterwordspacing

\bibitem{mqtt}
\BIBentryALTinterwordspacing
A.~Stanford-Clark and H.~L. Truong. (2014, Jan.) {MQTT} for sensor networks
  ({MQTT-SN}) protocol specification. [Online]. Available:
  \url{http://www.mqtt.org/new/wp-content/uploads/2009/06/MQTT-SN_spec_v1.2.pdf}
\BIBentrySTDinterwordspacing

\bibitem{johnsonLDPC}
S.~J. Johnson, ``Introducing low-density parity-check codes,'' \emph{University
  of Newcastle, Australia}, pp. 38--39, 2006.

\bibitem{11acSNR_Requirement}
\BIBentryALTinterwordspacing
B.~Verney. (2016, Oct.) 802.11n + 802.11ac data rates and {SNR} requirements.
  [Online]. Available:
  \url{https://higher-frequency.blogspot.com/2016/10/80211n-80211ac-data-rates-and-snr.html?m=1}
\BIBentrySTDinterwordspacing

\bibitem{Aruba_11ax_WP}
\BIBentryALTinterwordspacing
{Aruba Networks, Santa Clara, CA, USA}. (2018, Nov.) White paper 802.11ax.
  [Online]. Available:
  \url{https://www.arubanetworks.com/assets/wp/WP_802.11AX.pdf}
\BIBentrySTDinterwordspacing

\bibitem{Akbilek2019_11ax_Analysis}
A.~Akbilek, F.~Pfeiffer, and M.~Fuenfer, ``Analysis of {IEEE} 802.11 ax high
  efficiency {WLAN}s for in-vehicle use,'' in \emph{15th Wireless Congress:
  System and Applications}, 2018, pp. 1--9.

\bibitem{Adam}
D.~P. Kingma and J.~Ba, ``Adam: {A} method for stochastic optimization,'' in
  \emph{3rd International Conference on Learning Representations, {ICLR} 2015,
  San Diego, CA, USA, May 7-9, 2015, Conference Track Proceedings}, 2015.

\bibitem{RicHea:Learning-Based-Adaptive-Transmission:14}
A.~Rico-Alvari{\~n}o and R.~W. Heath, ``Learning-based adaptive transmission
  for limited feedback multiuser {MIMO}-{OFDM},'' \emph{IEEE Transactions on
  Wireless Communications}, vol.~13, no.~7, pp. 3806--3820, July 2014.

\end{thebibliography}
